\documentclass[ALICE,manyauthors]{cernphprep}

\usepackage[comma,square,numbers,sort&compress]{natbib}
\usepackage{hyperref}
\usepackage{lineno}
\usepackage{color}
\usepackage{units}
\usepackage{ulem}
\usepackage{amsmath}
\usepackage{soul}
\usepackage[T1]{fontenc}
\usepackage{orcidlink}
\usepackage{xspace}
\usepackage{color}
\usepackage{xcolor}
\usepackage{rotating}
\definecolor{light-gray}{gray}{0.8}

\newcommand{\pPb}{p--Pb\xspace}

\newcommand{\pt}{\ensuremath{{\it p}_{\rm T}}\xspace}

\newcommand{\cmnt}[1]{}

\newcommand{\GeVc}{\ensuremath{\mathrm{GeV}/c}\xspace}

\newcommand{\dph}{\Delta\varphi }

\begin{document}%
\begin{titlepage}
\PHyear{2023}
\PHnumber{024}      
\PHdate{27 February}  
%

\title{Azimuthal correlations of heavy-flavor hadron decay electrons with charged particles in pp and p--Pb collisions at $\pmb {\sqrt{s_{\rm{NN}}}}$ = \unit[5.02]{TeV}}
\ShortTitle{Heavy-flavor decay electron--charged particle correlations in pp, p$-$Pb collisions}   

\Collaboration{ALICE Collaboration\thanks{See Appendix~\ref{app:collab} for the list of collaboration members}}
\ShortAuthor{ALICE Collaboration} 

\begin{abstract}
The azimuthal ($\Delta\varphi$) correlation distributions between heavy-flavor decay electrons and associated charged particles are measured in pp and p--Pb collisions at $\sqrt{s_{\rm{NN}}} = 5.02$ TeV. Results are reported for electrons with transverse momentum $4<p_{\rm T}<16$ ${\rm GeV}/c$~and pseudorapidity $|\eta|<0.6$. The associated charged particles are selected with transverse momentum $1<p_{\rm T}<7$ ${\rm GeV}/c$,~and relative pseudorapidity separation with the leading electron $|\Delta\eta| < 1$. The correlation measurements are performed to study and characterize the fragmentation and hadronization of heavy quarks. The correlation structures are fitted with a constant and two von Mises functions to obtain the baseline and the near- and away-side peaks, respectively. The results from p--Pb collisions are compared with those from pp collisions to study the effects of cold nuclear matter. In the measured trigger electron and associated particle kinematic regions, the two collision systems give consistent results. The $\Delta\varphi$ distribution and the peak observables in pp and p--Pb collisions are compared with calculations from various Monte Carlo event generators.

\end{abstract}
\end{titlepage}
\setcounter{page}{2}

\section{Introduction}
\label{sec:Intro}

In high-energy hadronic collisions, heavy quarks (charm and beauty) are mainly produced in hard parton scattering processes. Due to the large momentum transfer characterizing these processes, their inclusive production cross sections can be calculated in the framework of perturbative quantum chromodynamics (pQCD)~\cite{Kniehl:2008zza,Cacciari:2003uh,Kniehl:2005mk,Cacciari:2003zu,ALICE:2022wpn}. The production cross sections of several open heavy-flavor hadrons and of their decay leptons in pp collisions were measured at both mid- and forward-rapidity at the LHC~\cite{ALICE:2019nuy,CMS:2017qjw,LHCb:2015swx,ALICE:2021wct,ALICE:2022mur,ALICE:2019nxm,ALICE:2019rmo,ALICE:2021mgk,ALICE:2012msv,ALICE:2012inj,ALICE:2012mzy,Aaij:2016jht,ALICE:2012acz,ALICE:2014ivb, Aaij:2012jd, Aad:2012jga, ATLAS:2013cia, Chatrchyan:2012hw, Khachatryan:2011mk, Chatrchyan:2011pw, Chatrchyan:2011vh, Khachatryan:2016csy}, and are described by pQCD  calculations~\cite{Kniehl:2012ti,Kniehl:2011bk,Cacciari:2012ny} with large theoretical uncertainties. The charm-hadron production cross section calculations in the pQCD frameworks are based on the factorization of parton distribution functions (PDF), the partonic cross section, and the fragmentation function. Recent measurements of charm-baryon production at midrapidity in pp collisions~\cite{ALICE:2022ych,ALICE:2022cop,ALICE:2021rzj,ALICE:2021npz,LHCb:2013xam,ALICE:2017thy,ALICE:2020wfu,ALICE:2020wla,Sirunyan:2019fnc,ALICE:2017dja,ALICE:2021psx,ALICE:2021bli} are not reproduced by pQCD calculations and event generators adopting a fragmentation model tuned on $\rm{e}^{+}\rm{e}^{-}$ data. A better description of these measurements can be obtained by models including hadronization mechanisms such as quark coalescence~\cite{Ravagli:2007xx}, additional color reconnections among parton fragments~\cite{Christiansen:2015yqa}, or by including enhanced feed-down from higher-mass charm-baryon states within a statistical hadronization approach~\cite{He:2019tik}, where the higher-mass excited charm-baryon states are predicted by the Relativistic Quark Model~\cite{Ebert:2011kk} but not yet measured. More differential measurements are needed to better understand the fragmentation (parton showering) and hadronization of heavy quarks. Two-particle angular correlations originating from heavy-flavor particles allow such processes to be characterized. 

The typical structure of a two-particle angular correlation distribution of high transverse-momentum ($p_{\rm{T}}$) trigger particles with associated charged particles features a “near-side” (NS) peak at $(\Delta\varphi, \Delta\eta) = (0,0)$ and an “away-side” (AS) peak at $\Delta\varphi = \pi$, extending over a wide pseudorapidity range. The NS peak is mainly induced by particles emerging from the fragmentation of the same parton that produced the trigger particle. The AS peak is related to the fragmentation of the other parton produced in the hard scattering. Here, $\Delta\eta$ is the difference in pseudorapidity between the trigger and associated particles. The peaks lie on top of an approximately flat continuum extending over the full $(\Delta\varphi, \Delta\eta)$ range~\cite{ALICE:2016clc}. At leading order (LO) accuracy in QCD, heavy quark--antiquark pairs are produced back-to-back in azimuth~\cite{Mangano:1991jk}.  At next-to-leading order (NLO), the correlation shapes can significantly differ from such a topology~\cite{Mangano:1991jk, Norrbin:2000zc}. Gluon radiation of heavy quarks can smear the back-to-back topology and broaden the near- and away-side peaks. In the gluon splitting process, the two heavy quarks can be produced with a small opening angle, depending on the $p_{\rm{T}}$ of the gluon and the mass of the produced quark, generating two sprays of hadrons that can partially overlap, leading to a broader near-side peak. In the flavor excitation process~\cite{Norrbin:2000zc}, the heavy-quark pairs can be significantly separated in rapidity, and the hadrons from the opposite quark with respect to the trigger particle induce a nearly flat contribution to the $\Delta\varphi$ distribution. The correlation measurements provide insight into heavy-flavor jet properties at low transverse momentum. By varying the $p_{\rm{T}}$ interval of the trigger and associated particles, the correlation measurements allow the details of jet fragmentation to be studied, such as the jet angular profile and the momentum distribution of the particles produced in the fragmentation of the hard parton.

The azimuthal correlation distributions of prompt D mesons and charged particles were measured by the ALICE Collaboration in pp collisions at $\sqrt{s} = 5.02,  7,$ and 13 TeV for $p_{\rm T}^{\rm D}$ of the D~mesons up to 36 \GeVc ~and associated charged particles up to $p_{\rm{T}}^{\rm{assoc}} =$  3 \GeVc ~\cite{ALICE:2016clc, ALICE:2019oyn, ALICE:2021kpy}. The measurements were compared with Monte Carlo (MC) simulations with
different event generators, like PYTHIA~\cite{Sjostrand:2006za, Sjostrand:2007gs,Singh:2023zuz}, HERWIG~\cite{Kersevan:2004yg, Bellm:2015jjp}, EPOS~\cite{Werner:2010aa,Drescher:2000ha}, and POWHEG coupled with PYTHIA8 for the parton shower and hadronization (POWHEG+PYTHIA8)~\cite{Nason:2004rx,Frixione:2007vw}. A substantial difference among the generators was observed, with PYTHIA8 and POWHEG+PYTHIA8 providing the best description of the measured observables. These differences can be ascribed to the specific implementation of features such as hard-parton scattering matrix elements, parton showering, hadronization algorithm, and underlying event generation, affecting the correlation functions of heavy-flavor hadrons and charged particles. 
Measuring the correlation distribution between heavy-flavor decay electrons and charged particles grants a substantially larger sample of correlation pairs, compared to measurements of D mesons and charged particle azimuthal correlations~\cite{ALICE:2019oyn, ALICE:2016clc}. This allows a significant extension of the $p_{\rm T}^{\rm assoc}$ range of associated particles and can provide a more complete picture of the heavy quark fragmentation. In addition, electrons originating from beauty-hadron decays (b $\rightarrow$ (c $\rightarrow$) e) dominate the heavy-flavor hadron decay electron spectrum ($> 50\%$) at high $p_{\rm T}^{\rm e}$~($>5$ \GeVc)~\cite{ALICE:2014aev}. Hence, probing large enough trigger electron transverse momenta enables the study of the correlation function of particles originating from beauty-hadron decays, and provides information on the different correlation structures for charm and beauty quarks. This additional information can be used to further constrain the MC simulations. 
These advantages come at the price of an additional smearing introduced in the correlation function, due to the non-zero angle between the trigger electron direction and the direction of the parent heavy-flavor hadron before its decay. The momentum of the electron could also be further away from the quark momentum as compared to that of the parent hadron due to its decay kinematics. 

In proton--nucleus (p--A) collisions, several cold nuclear-matter effects can influence the production, fragmentation, and hadronization of heavy quarks~\cite{ALICE:2022wpn}. In the initial state, the parton distribution functions (PDFs) are modified in bound nucleons as compared to free nucleons. This feature is described by phenomenological parameterizations referred to as nuclear PDFs (nPDFs)~\cite{Eskola:2009uj,deFlorian:2003qf,Hirai:2007sx}. When the production process is dominated by gluons at low Bjorken-$x$, the nucleus can be described by the Color-Glass Condensate (CGC) effective theory as a coherent and saturated gluonic system~\cite{Fujii:2013yja,Tribedy:2011aa,Albacete:2012xq,Rezaeian:2012ye}. The CGC predicts momentum correlations in the initial state, that would impact the angular correlations of the produced heavy-quark pairs. 
Partons can also undergo multiple elastic, inelastic, and coherent scatterings, due to the presence of the nucleus in the initial state~\cite{Accardi:2009qv,Salgado:2011wc} and to possible parton interactions in the high-density environment in the final state, particularly in collisions with large charged-particle multiplicity. These effects can be studied by measuring modifications in the angular shape or in the associated-particle peak yields of the angular correlation distributions of heavy-flavor particles with charged hadrons~\cite{ALICE:2016clc,ALICE:2019oyn}. Measurements of azimuthal correlations of prompt D~mesons and charged hadrons in p--Pb collisions by the ALICE collaboration~\cite{ALICE:2016clc,ALICE:2019oyn}, showed that the near- and away-side peaks of the correlation distribution are consistent with those measured in pp collisions in the same kinematic region. Employing heavy-flavor decay electrons as trigger particles in place of prompt D mesons allows studying the impact of cold-nuclear-matter effects for a wider associated particle $p_{\rm T}^{\rm assoc}$ range, as well as to investigate their impact on the beauty-quark fragmentation and hadronization. 

In heavy-ion collisions, a strongly-interacting matter consisting of deconfined quarks and gluons, the quark$-$gluon plasma (QGP), is produced~\cite{PHENIX:2004vcz, STAR:2005gfr,PHOBOS:2004zne,BRAHMS:2004adc,ALICE:2010yje,ALICE:2022wpn}. In the presence of the QGP, high-$p_{\rm{T}}$ partons lose energy via medium-induced gluon radiation and collisions with the medium constituents~\cite{ALICE:2021rxa,ALICE:2021kfc,Gyulassy:1990ye,Baier:1996sk,Thoma:1990fm,Braaten:1991we}. These interactions cause a modification of the heavy-quark fragmentation and induce a broadening of the emerging jets and a softening of their constituents~\cite{Borghini:2009eq,Connors:2017ptx}. Two-particle angular correlations have been extensively used to search for remnants of the radiated energy and to probe the medium response to the high-$p_{\rm{T}}$ parton. The recent measurement of angular correlations between D mesons and charged particles in Au--Au collisions by the STAR Collaboration~\cite{STAR:2019qbf}, shows a significant modification of the near-side peak width and associated yield, which increases from peripheral to central collisions. Measurements of angular correlations between electrons from heavy-flavor hadron decays and charged particles by the PHENIX Collaboration show modifications of the away-side peak yield and width in Au--Au collisions compared to pp collisions~\cite{PHENIX:2010cfl}. For future studies of heavy-flavor hadron correlations in heavy-ion collisions at the LHC, similar measurements in pp and p--Pb collisions are crucial to serve as reference~\cite{ALICE:2022wwr}. 

In this article, ALICE measurements of the azimuthal correlations between electrons from heavy-flavor hadron decays with associated charged particles in pp collisions at center-of-mass energy $\sqrt{s}=5.02$ TeV and p--Pb collisions at center-of-mass energy per nucleon--nucleon collision $\sqrt{s_{\rm NN}} = 5.02$ TeV are reported. The correlation distributions are measured for trigger electrons originating from heavy-flavor hadron decays in the $p_{\rm T}^{\rm e}$ range $4 < p_{\rm T}^{\rm e} < 12$ \GeVc ~and associated charged particles in the range $1 < p_{\rm T}^{\rm assoc} < 7$ \GeVc, the latter granting a significantly higher $p_{\rm T}^{\rm assoc}$ reach compared to previously published correlation measurements of D mesons with charged particles~\cite{ALICE:2019oyn, ALICE:2016clc}. The correlation distributions for trigger electron $p_{\rm T}^{\rm e}$ in the range $4 < p_{\rm T}^{\rm e} < 7$ \GeVc ~and $7 < p_{\rm T}^{\rm e} < 16$ \GeVc ~are also measured in order to study correlation shapes in kinematic ranges where the electrons are dominantly produced by charm- and beauty-hadron decays, respectively.

 The article is organized as follows - in Sec.~\ref{sec:ALICE}, the ALICE apparatus, its main detectors used in the analyses, and the data samples are reported. The complete analysis procedure is described in Sec.~\ref{sec:Analysis}. The systematic uncertainties associated with the measurements are discussed in Sec.~\ref{sec:Systematics}. The analysis results are presented and discussed in Sec.~\ref{sec:Results}. The article is briefly summarized in Sec.~\ref{sec:Summary}.

\section{Experimental apparatus and data samples}
\label{sec:ALICE}

The ALICE apparatus consists of a central barrel, covering the pseudorapidity region $|\eta| < 0.9$,
a muon spectrometer with $-4 < \eta < -2.5$ coverage, and forward- and backward-pseudorapidity detectors employed for triggering, background rejection, and event characterization. A complete description of the
detector and an overview of its performance are presented in Refs.~\cite{ALICE:2014sbx,Aamodt:2008zz}. The central-barrel detectors used in the analysis are the Inner Tracking System (ITS), the Time Projection Chamber (TPC), and the electromagnetic calorimeters (EMCal and DCal). They are embedded in a large solenoidal magnet that provides a maximum magnetic field of $B = 0.5$ T parallel to the beam direction. The ITS~\cite{ALICE:2010tia} consists of six layers of silicon detectors, with the innermost two composed of Silicon Pixel Detectors (SPD). 
The ITS was used to reconstruct the primary vertex and the charged particle tracks. 
The TPC~\cite{Alme:2010ke} is a gaseous chamber capable of three-dimensional reconstruction of charged-particle tracks, and is the main tracking detector of the central barrel. Moreover, it enables charged-particle identification via the measurement of the particle specific energy loss (d$E/$d$x$) in the detector gas. 
 The EMCal and DCal detectors~\cite{Cortese:1121574, Allen:2010stl} are shashlik-type sampling calorimeters consisting of alternate layers of lead absorber and scintillator material. The EMCal covers ranges of $|\eta| < 0.7$ in pseudorapidity and $\Delta\varphi = 107^{\circ}$ ($ 80^{\circ} < \varphi < 187^{\circ}$) in azimuth. The DCal is located azimuthally opposite the EMCal, with a coverage of $0.22 < |\eta| < 0.7$ and $\Delta\varphi = 60^{\circ}$ ($ 260^{\circ} < \varphi < 320^{\circ}$) and $|\eta| < 0.7$ and $\Delta\varphi = 7^{\circ}$ ($ 320^{\circ} < \varphi < 327^{\circ}$). For the remaining part of this article, EMCal and DCal will be together referred to as EMCal, as they are part of the same detector system, used for electron identification. Two scintillator arrays, the V0 detector~\cite{ALICE:2013axi}, placed on each side of the interaction point (with pseudorapidity coverage $2.8 < \eta < 5.1$ and $-3.7 < \eta < -1.7$) were utilized for triggering and offline rejection of beam-induced background events. The minimum bias trigger was defined requiring coincident signals in both scintillator arrays of the V0 detector.
In \pPb collisions, the contamination from beam-induced background interactions and electromagnetic interactions was further removed with the information of the Zero Degree Calorimeters (ZDC)~\cite{Arnaldi:1999zz}, located along the beam line at 112.5 m on both sides of the interaction point. A T0 detector~\cite{ALICE:2016ovj}, composed of two arrays of quartz Cherenkov counters, covering an acceptance of $4.6 < \eta < 4.9$ and $-3.3 < \eta < -3.0$, was employed to determine the luminosity together with the V0 detector.

The results presented in this paper were obtained using minimum bias triggered data recorded with the ALICE detectors during the LHC Run~2 
from pp collisions at $\sqrt{s}=5.02$ TeV and
from \pPb collisions at $\sqrt{s_{\rm{NN}}}=5.02$ TeV. Pile-up events containing two or more primary vertices were rejected using an algorithm based on the detection of multiple vertices reconstructed from track segments in the SPD.
In order to obtain a uniform acceptance of the detectors, only events with a reconstructed primary vertex within $\pm10$ cm from the center of the detector along the beam line were considered for both pp and p--Pb collisions. The number of selected pp and p--Pb events are about 800M and 546M, respectively, corresponding to integrated luminosities of $(16.63 \pm 0.32)$ nb$^{-1}$~\cite{ALICE:2018pqt} and $(250 \pm 10)$ $\rm{\mu b^{-1}}$~\cite{ALICE:2014gvw}.

\section{Analysis overview}
\label{sec:Analysis}

The measurements of two-particle azimuthal correlations between electrons from heavy-flavor hadron decays (trigger) and charged (associated) particles were obtained from the correlation distributions of all identified electrons after subtracting the contributions which do not originate from heavy-flavor hadron decays. Effects from the limited two-particle acceptance and detector inhomogeneities were corrected using the event-mixing technique. 
The per-trigger correlation distributions were corrected for the associated-particle reconstruction efficiency. They were not corrected for the trigger-electron efficiency, as the efficiency was found to be \pt independent, and the correction factor would cancel with the per-trigger normalization. The properties of the correlation distribution in $\dph$, peak yields and widths, were obtained by applying a fit to the corrected $\Delta\varphi$ distribution. A detailed description of the above mentioned analysis procedures is provided in the following sections. The analysis technique is the same in both pp and p--Pb measurements (unless specified otherwise in the text).
Throughout this paper, the term ``electron" refers to both electrons and positrons. 

\subsection{Electron identification and associated-particle reconstruction}
Electrons with transverse momentum in the interval $4 < p_{\rm T}^{\rm e} < 16$~\GeVc ~and $|\eta| < $ 0.6 were selected using similar criteria as those discussed in Ref.~\cite{ALICE:2019nuy}. Tracks were required to have at least one hit in any of the two SPD layers in order to reduce the contamination of electrons from photon conversions in the detector material. In order to reject secondary electrons~\cite{ALICE-PUBLIC-2017-005}, produced in interactions with the detector material or from weak decays of long-lived particles, the tracks were required to have a distance of closest approach to the primary vertex of less than 1 cm along the beam axis and 0.5 cm in the transverse plane. To ensure the selection of high-quality tracks, electron tracks were required to have a minimum of 70 crossed pad rows in the TPC (out of 159) and a minimum fraction of 0.8 of found space points relative to the maximum value, driven by the track direction~\cite{ALICE:2020jsh}.
The particle identification employed a selection on d$E/$d$x$ inside the TPC and on the energy deposited in the EMCal detector. The discriminant variable used for the TPC detector is the deviation of d$E/$d$x$ from the parameterized Bethe--Bloch expectation value for electrons~\cite{Bethe:1930ku}, expressed in terms of d$E/$d$x$ resolution, $n\sigma^{\rm{TPC}}_{\rm{e}}$. An asymmetric selection of  $-1 < n\sigma^{\rm{TPC}}_{\rm{e}} < 3$ was applied as the background contamination is higher for negative $n\sigma^{\rm{TPC}}_{\rm{e}}$. Additionally, electrons were identified and separated from hadrons using the $E/p$ information from the EMCal detector, where $E$ is the energy of the EMCal cluster (deposited by the particle while crossing the detector)~\cite{ALICE:2022qhn, ALICE:2019nuy}, and $p$ is the momentum of the track measured by the TPC, along with a condition on the elliptical shape of the EMCal cluster, $\sigma_{\rm{long}}^{2}$~\cite{ALICE:2022qhn}. The electron sample was obtained by selecting candidates with $0.8 < E/p < 1.2$, as expected for electrons, while hadrons have lower $E/p$ values, and with $0.02 < \sigma_{\rm{long}}^{2} < 0.9$. The lower threshold on $\sigma_{\rm{long}}^{2}$ removes contamination caused by neutrons hitting the readout electronics. 

Associated particles were defined as all charged primary particles~\cite{ALICE-PUBLIC-2017-005} with pseudorapidity $|\eta| < 0.8$ and $\pt > 1$~\GeVc. Reconstructed tracks were required to have a minimum of 60 crossed pad rows in the TPC (out of 159) and a minimum fraction of 0.6 of found space points relative to the expected maximum considering the track position in the detector geometry~\cite{ALICE:2020jsh}. Additional requirements on the distance of closest approach to the primary vertex of less than 1 cm along the beam axis and 0.5 cm in the transverse plane were applied. 
The associated particles were also required to have a \pt smaller than the trigger electron \pt. 
This condition induces a kinematic bias for the regions where the trigger and associated \pt ranges overlap, that can be reproduced by simulations and model predictions.

\subsection{Azimuthal correlation distribution and mixed-event correction}\label{sec:MEcorrection}
The two-dimensional correlation distribution as a function of azimuthal angle difference ($\Delta\varphi = \varphi_{\rm{e}} - \varphi_{\rm{ch}}$) and pseudorapidity difference ($\Delta\eta = \eta_{\rm{e}} - \eta_{\rm{ch}}$) between electron and charged particles, $C(\Delta\varphi, \Delta\eta)$, was computed for the \pt interval $4 < p_{\rm T}^{\rm e} < 12$~\GeVc, as well as the two intervals $4 < p_{\rm T}^{\rm e} < 7$~\GeVc ~and $7 < p_{\rm T}^{\rm e} < 16$~\GeVc, and for five \pt intervals of associated particles between 1 and 7~\GeVc ~($1 < p_{\rm T}^{\rm assoc} < 2$~\GeVc,  $2 < p_{\rm T}^{\rm assoc} < 3$~\GeVc, $3 < p_{\rm T}^{\rm assoc} < 4$~\GeVc, $4 < p_{\rm T}^{\rm assoc} < 5$~\GeVc, and $5 < p_{\rm T}^{\rm assoc} < 7$~\GeVc). For each kinematic interval, the correlation distributions were corrected for the limited pair acceptance and for the detector inhomogeneities using the event-mixing technique~\cite{ALICE:2013snk} as shown in Fig.~\ref{fig:MEcorrectionpp} in Appendix~\ref{SuppleFigs}. The mixed-event correlation distribution, $ME(\Delta\varphi, \Delta\eta)$, was obtained by correlating electrons in an event with charged particles from other events with similar multiplicity and primary-vertex position along the beam direction. The distribution obtained from the mixed events features a triangular-like shape as a function of $\Delta\eta$, due to the limited $\eta$ coverage of the detector, and is approximately flat as a function of $\Delta\varphi$. Any non-flatness in $\Delta\varphi$ would be due to $\varphi$-dependent detector inefficiencies and inhomogeneities. 
At $(\Delta\varphi, \Delta\eta) \approx (0,0)$, the trigger and associated particle experience the same detector effects and the per-trigger correlation distribution is thus not affected. This property can be used to obtain the normalization factor, $\beta$, for the mixed event distribution, defined as the average number of counts in the range $-0.2 < \Delta\varphi < 0.2$ and $-0.07 < \Delta\eta < 0.07$. 

The mixed-event corrected correlation distribution, d$^2N/($d$\Delta\eta$d$\Delta\varphi$), labeled as $S(\Delta\eta, \Delta\varphi)$, was obtained as the ratio of the correlation distribution from the same event to the mixed event distribution, scaled by $\beta$, i.e.,

\begin{equation}
\frac{{\rm d}^{2}N}{{\rm d}\Delta\eta {\rm d}\Delta\varphi} \equiv S(\Delta\eta, \Delta\varphi) = \beta \times \frac {C (\Delta\eta, \Delta\varphi)} {ME(\Delta\eta, \Delta\varphi)}.
\end{equation}

The two-dimensional correlation distribution was subject to significant statistical fluctuations, due to the limited size of the heavy-flavor decay electron sample, especially at large $|\Delta\eta|$ values. To grant larger precision to the results, the mixed-event corrected azimuthal correlation distribution was integrated over in the range $|\Delta\eta| < 1$ to obtain a one-dimensional $S(\Delta\varphi)$ distribution.

\subsection{Background subtraction}
\label{BackSub}
The hadron contamination in the selected electron sample was estimated by considering tracks identified as hadrons using $n\sigma^{\rm{TPC}}_{\rm{e}}<-3.5$. The $E/p$ distribution of hadrons was scaled to match the electron-candidate $E/p$ distribution in the interval $0.3 < E/p < 0.65$, away from the electron signal region, similar to the procedure discussed in Ref.~\cite{ALICE:2019bfx}. The contamination from charged hadrons was estimated to be around $1\%$ at \pt = 4~\GeVc ~increasing to about $12\%$  at 16~\GeVc ~in both pp and p--Pb collisions. The hadron contamination in the azimuthal distribution of the inclusive electron sample was obtained using the correlation distributions of trigger particles with $n\sigma^{\rm{TPC}}_{\rm{e}} < -3.5 $, which was scaled to match the estimated hadron contamination. It was then subtracted from the inclusive electron (InclE) correlation distribution.

The selected electrons are composed of signal electrons originating from heavy-flavor hadron decays (HFe),  
and background electrons. The main background source is constituted by Dalitz decays of neutral mesons ($\pi^0$ and $\eta$) and photon conversions in the detector material, which produce electron--positron pairs with low invariant mass, peaked around zero.  
 The background electrons were identified using an invariant-mass technique~\cite{ALICE:2015zhm, ALICE:2016mpw}, where each selected electron was combined with oppositely-charged partner electrons, obtaining unlike-sign (ULS) pairs and calculating their invariant mass ($ m_{\rm e^{+}e^{-}}$). The partner electrons were selected by applying similar but looser track-quality and particle-identification criteria than those used for selecting the signal electrons, in order to increase the efficiency of finding the partner~\cite{ALICE:2016mpw,ALICE:2018yau}. Electron--positron pairs from the  
 background have a small invariant mass, while random combinations including heavy-flavor decay electrons forming a pair with other electrons gives a wider invariant-mass distribution. This combinatorial contribution was estimated from the invariant-mass distribution of like-sign electron (LS) pairs. The $S(\dph)$ distributions of electrons composing ULS and LS pairs, $S(\dph)^{\rm{ULS}}$ and $S(\dph)^{\rm{LS}}$, respectively, were obtained. The 
background contribution was then evaluated by subtracting the LS distribution from the ULS distribution in the invariant mass region $m_{\rm e^{+}e^{-}} < 0.14$~\GeVc. 
The efficiency of finding the partner electron, referred to as the tagging efficiency ($\varepsilon_{\rm tag}$) from here on, was estimated using MC simulations. In the pp and \pPb analyses, the MC sample was obtained using PYTHIA~6.4.25 event generator~\cite{Sjostrand:2006za}, with the Perugia 2011 tune~\cite{Skands:2010ak}, and HIJING~1.36~\cite{Wang:1991hta} generators, respectively. They will be referred to as PYTHIA6 and HIJING in the following. The generated particles in all MC samples were propagated through the ALICE apparatus using GEANT 3.21.11~\cite{Brun:1073159}. In order to increase the statistical precision of the tagging efficiency, $\pi^0$ and $\eta$ meson samples with a flat \pt shape, generated with PYTHIA6,  were embedded in the simulated events. The biased \pt shape was corrected by applying a weight to reproduce the measured \pt~spectra as described in~\cite{ALICE:2019bfx, ALICE:2020try}. The tagging efficiency for pp (p--Pb) collisions was about $74\%$ ($75\%$) at \pt = 4~\GeVc, increasing to about $79\%$ ($77\%$) for $\pt > 7$~\GeVc. The $\dph$ correlation distribution of background electrons was corrected by the tagging efficiency and subtracted from the inclusive electron distribution, that was already corrected for the hadron contamination, to obtain the azimuthal distribution of electrons from heavy-flavor hadron decays ($S(\dph)^{\rm{HFe}}$), 

\begin{equation}
    S(\dph)^{\rm{HFe}} = S(\dph)^{\rm{InclE}}  -  \frac{1}{\varepsilon_{\rm{tag}}} [S(\dph)^{\rm{ULS}} - S(\dph)^{\rm{LS}}]
    \label{eg:hfe}.
\end{equation}

Contributions from other sources, such as decays of $J/\psi$ and kaons, are negligible in the \pt~ranges considered in this analysis~\cite{ALICE:2015zhm}.

The azimuthal correlation distribution of electrons from heavy-flavor hadron decays and charged particles has to be corrected for the inefficiencies in the reconstruction of the associated particles and for the contamination of secondary particles in the associated particle sample. The reconstruction efficiency for charged primary particles was obtained using a different MC sample without any embedded particles using PYTHIA6~\cite{Sjostrand:2006za} and HIJING~\cite{Wang:1991hta} generators for pp and p--Pb collisions, respectively. The efficiency obtained was in the range 86--90\% (85--92\%) in the $1 < \pt < 7$~\GeVc interval ~for pp (p--Pb) collisions.  

The amount of contamination from secondary particles~\cite{ALICE-PUBLIC-2017-005} was also estimated using the same MC simulations, and shows values in the range 2--4\% in pp collisions and 4--6\% in p--Pb collisions, for the \pt interval considered.  The fully-corrected azimuthal-correlation distribution was divided by the number of electrons originating from heavy-flavor hadron decays ($N_{\rm (c, b) \rightarrow e}$), to obtain a per-trigger normalization, where $N_{\rm (c, b) \rightarrow e}$ is expressed as
\begin{equation}
    N^{\rm (c, b) \rightarrow e} = N_{\rm{InclE}} - \frac{1}{\varepsilon_{\rm{tag}}} [N_{\rm{ULS}} - N_{\rm{LS}}]
    \label{eg:Nhfe}.
\end{equation}

\subsection{Characterization of the azimuthal distribution}
In order to quantify the properties of the measured azimuthal correlation, the following fit function was used

\begin{equation}
    f(\Delta \varphi) = b + Y_{\mathrm{NS}} \frac{e^{\kappa_{\rm NS}\cos{(\Delta \varphi)}}}{2{\pi}I_{0}(\kappa_{\rm NS})} + Y_{\mathrm{AS}} \frac{e^{\kappa_{\rm AS}\cos{(\Delta \varphi - \pi)}}}{2{\pi}I_{0}(\kappa_{\rm AS})}.
    \label{eqvon}
\end{equation}

It is composed of two von Mises functions, to model circular data, describing the NS and AS peaks, and a constant term, $b$, describing the baseline, which is a free parameter. 
The terms $\kappa_{NS}$ and $\kappa_{AS}$ in the von Mises function are the  measure of concentration of NS and AS peak, respectively, where $1/\kappa$ is analogous to the variance $\sigma^{2}$, and $I_{0}$ is the zeroth-order modified Bessel function evaluated at $\kappa$. The parameters $Y_{\mathrm{NS}}$ and $Y_{\mathrm{AS}}$ represent the integral of the near- and away-side peaks, respectively. By symmetry considerations, the means of the NS and AS peaks are fixed to $\dph = 0$ and $\dph = \pi$, respectively. The baseline $b$ represents the physical minimum of the $\dph$ distribution. The width ($\sigma$) of the peaks is given by 

\begin{align}
\sigma = \sqrt{-2\log\frac{I_{1}(\kappa)}{{I_{0}(\kappa)}}, }
\label{eqsigma}
\end{align}

where $I_{1}$ is the first-order modified Bessel function evaluated at $\kappa$. The per-trigger yields of the NS and AS peaks were obtained by integrating the bin counts in the ranges $-3\sigma_{\rm{NS}} < \dph < 3\sigma_{\rm{NS}}$ and $-3\sigma_{\rm{AS}} < \dph-\pi < 3\sigma_{\rm{AS}}$, respectively, after subtracting the baseline value $b$ from the distribution.

\section{Systematic uncertainties}
\label{sec:Systematics}

The $\dph$ correlation distribution and the per-trigger NS and AS yields and widths are affected by systematic uncertainties, related to the procedures used for electron-track selection,  identification and subtraction of the hadron contamination, background-electron subtraction, associated-particle efficiency correction, mixed-event correction, and fitting routine applied to the correlation distribution. The uncertainties from each of these sources were estimated separately, by varying the selection criteria or by using an alternative approach to the one described in the previous section. For each variation, its effect on the NS and AS peak yields and widths was obtained by reevaluating these observables after fitting and subtracting the baseline of the resulting correlation distribution. The uncertainties were computed separately for each trigger electron and associated particle $\pt$ range. The systematic uncertainties on the correlation distribution from associated-particle efficiency correction and mixed-event correction are considered as correlated in $\dph$. The remaining sources are considered as uncorrelated in $\dph$. A summary of the systematic uncertainties of the correlation distribution, NS and AS yields and widths for $4 < \pt^{\rm e} < 12$~\GeVc ~are reported in Tables~\ref{tab:SysCombined_pp_pte412} and~\ref{tab:SysCombined_pPb_pte412} for pp and p--Pb collisions, respectively. The $\dph$ correlated and uncorrelated uncertainties are separately reported for the $\dph$ distribution, and the total uncertainty from all sources is reported for the peak yields and widths.

Possible biases related to the specific track quality selection for electrons used in the analysis were studied by varying the selection criteria~\cite{ALICE:2019nuy}. An uncertainty of 1--2\% on the correlation distribution was obtained as a function of $\pt^{\rm{assoc}}$ for  $4 < \pt^{\rm e} < 12$~\GeVc in both collision systems. For the NS and AS yields, an uncertainty in the range 1--2\% was estimated. The uncertainty from track selection on the NS and AS widths was found to be negligible. 

The uncertainty due to the electron identification using the TPC and EMCAL signals was estimated by varying the selection criteria for $n\sigma^{\rm{TPC}}_{\rm{e}}$, $E/p$, and  $\sigma_{\rm{long}}^{2}$. The chosen variations change the efficiency by a maximum of $\sim$ 20\%. A total uncertainty from these sources of 2--5\% was obtained for the correlation distribution as a function of $\pt^{\rm{assoc}}$ in pp and p--Pb collisions, for $4 < \pt^{\rm e} < 12$~\GeVc. The resulting uncertainties ranged between 2\% and 6\% for the NS and AS yields, and between 2\% and 7\% for the NS and AS widths.

The contribution from background electrons was estimated using the invariant-mass method. The systematic uncertainty of the procedure, mainly affecting the average tagging efficiency,
was obtained by varying the selection criteria of the partner electron tracks, including the minimum \pt and the invariant-mass window of the electron--positron pairs. The variation affects the tagging efficiency by $\sim$ 5\%. A resulting systematic uncertainty of 1--2\% was obtained as a function of $\pt^{\rm{assoc}}$ on the correlation distribution, the peak yields, and their widths for $4 < \pt^{\rm e} < 12$~\GeVc in pp and p--Pb collisions.

The uncertainty related to the specific selection of associated particles was estimated by varying the charged track selection criteria, including a requirement of a hit in one of the two SPD layers of the ITS, and varying the selection on the distance of closest approach, which affects the secondary particle contamination. This uncertainty is considered correlated in $\dph$. For $4 < \pt^{\rm e} < 12$~\GeVc, uncertainties of 1--2\% and 2--3\% were obtained for the correlation distribution in pp and p--Pb collisions, respectively. For NS and AS yields, an uncertainty of 1--3\% and 1--4\% was estimated for pp and p--Pb collisions, respectively. Uncertainties of less than 3\% and 4\% were obtained for the NS and AS widths in pp and p--Pb collisions, respectively.

Effects induced by the limited detector acceptance and its local inhomogeneities were corrected using the mixed-event technique. The normalization factor, $\beta$, was varied by taking the integrated yield over the full $\dph$ range for $|\Delta\eta| < 0.01$. A correlated uncertainty in $\dph$ of 1\% was obtained for the correlation distribution and the peak yields in pp and p--Pb collisions, respectively. No uncertainty was assigned for the NS and AS widths. 

The $\dph$ distribution can be affected in case of a non-zero $v_2$ of HFe and charged particles. As there are no previous measurements of HFe $v_2$ in minimum bias pp and p--Pb collisions, a conservative estimate was obtained using the measurements in 0--20\% central p--Pb collisions in Ref.~\cite{ALICE:2018gyx}. The inclusion of $v_2$ has an impact of less than 1\% on the baseline and peak yields, and does not modify the NS and AS widths.

Several checks were performed to study the stability of the fit to the correlation distributions.
Alternative functions, i.e., a Gaussian and a generalized Gaussian, were used to fit the NS and AS peaks instead of the von Mises function. Alternative fits were also performed fixing the baseline value to the average of the points in the transverse region, defined as $\pi/3 < |\dph| < \pi/2$, to study its stability given statistical fluctuations.
In place of the default bin counting procedure, the NS and AS yields were obtained as the integral of the fit functions in the range $-3\sigma_{\rm{NS}} < \dph < 3\sigma_{\rm{NS}}$ and $-3\sigma_{\rm{AS}} < \dph-\pi < 3\sigma_{\rm{AS}}$. The overall systematic uncertainty was calculated by taking the maximum variation of the results. The uncertainty from the baseline estimation on the correlation distribution is quoted as absolute numbers affecting all $\dph$ bins by the same value. The uncertainty of the NS and AS yields and width varies in the range 4--9\% and 10--11\% for pp and p--Pb collisions, respectively, for $4 < \pt^{\rm e} < 12$~\GeVc.  

Similar procedures were followed to estimate the systematic uncertainties from the above mentioned sources on the correlation distribution, NS and AS yields and widths for $4 < \pt^{\rm e} < 7$~\GeVc ~and $7 < \pt^{\rm e} < 16$~\GeVc. The uncertainty values were found to be similar to those obtained for $4 < \pt^{\rm e} < 12$~\GeVc ~in both collision systems. 

\begin{table}[tbp]
\centering
\caption{Systematic uncertainties of the correlation distribution, the peak yields, and their widths for $4 < \pt^{\rm e} < 12$~\GeVc in pp collisions. The individual sources of systematic uncertainties depend on the associated particle $\pt$. The values presented as a range correspond to the lowest and highest $\pt^{\rm{assoc}}$~interval. 
For the correlation distribution, the systematic uncertainty from the baseline estimation is given as absolute value, and the total uncertainties from correlated and uncorrelated sources are reported separately.
}
\begin{tabular}{ l c c c c c}
 \hline
 Source & Correlation distribution & NS yield & AS yield & NS width & AS width \\
 \hline
 Electron track selection & 1\% & 1\% & 1\% & 0\% & 0\% \\
 Electron identification & 3--5\% & 2--4\% & 3--6\% & 2--6\% & 4--7\% \\
 Background electron & 1\% & 1\% & 2\% & 1\% & 1\%  \\
 Associated particle selection & 1--2\% & 1--2\% & 1--3\% & 1--3\% & 1--3\%\\
 Mixed-event correction & 1\% & 1\% & 1\% & 0\% & 0\% \\
 Fit routine / Baseline estimation & 0.001--0.02 ($\rm{rad}^{-1}$) & 5--8\% & 8--9\% & 10\% & 10\% \\
  \hline
 
 Total (correlated sources) & 1--2\% &  &  &  &  \\ 
 Total (uncorrelated sources) & 3--5\% &  & & &  \\ 
 Total &  & 6--9\% & 9--11\% & 10--12\% & 11--13\% \\ 
  \hline
\end{tabular}
\label{tab:SysCombined_pp_pte412}
\end{table}

\begin{table}[tbp]
\centering
\caption{Systematic uncertainties of the correlation distribution, the peak yields, and their widths for $4 < \pt^{\rm e} < 12$~\GeVc in p--Pb collisions. The individual sources of systematic uncertainties depend on the associated particle $\pt$. The values presented as a range correspond to the lowest and highest $\pt^{\rm{assoc}}$~interval. The systematic uncertainty of the correlation distribution from the baseline estimation is given as absolute values.  For the correlation distribution, the systematic uncertainty from the baseline estimation is given as absolute value, and the total uncertainties from correlated and uncorrelated sources are reported separately.}
\begin{tabular}{ l c c c c c} 
 \hline
 Source & Correlation distribution & NS yield & AS yield & NS width & AS width \\
 \hline
 Electron track selection & 1--2\% & 1\% & 1\% & 1\% & 1\% \\
 Electron identification & 2--4\% & 4\% & 4\% & 2--4\% & 4--5\% \\
 Background electron & 1\% & 1\% & 1\% & 1\% & 1\%  \\
 Associated particle selection & 2--3\% & 2--4\% & 2--4\% & 1--4\% & 2\%\\
 Mixed-event correction & 1\% & 1\% & 1\% & 0\% & 0\%  \\
 Fit routine / Baseline estimation &  0.0005--0.02 ($\rm{rad}^{-1}$) & 4--5\% & 6--7\% & 11\% & 11\% \\
  \hline

  Total (correlated sources) & 2--3\% &  & &  &  \\
 Total (uncorrelated sources) & 2--5\% &  &  &  &  \\
Total &  & 6--8\% & 8--9\% & 11--13\% & 12\%  \\
  \hline
\end{tabular}
\label{tab:SysCombined_pPb_pte412}
\end{table}

\newpage

\section{Results}
\label{sec:Results}

\subsection{Comparison of the results in pp and p--Pb collisions}
\begin{figure}[!h]
\centering
\subfigure{
\includegraphics[scale=0.8]{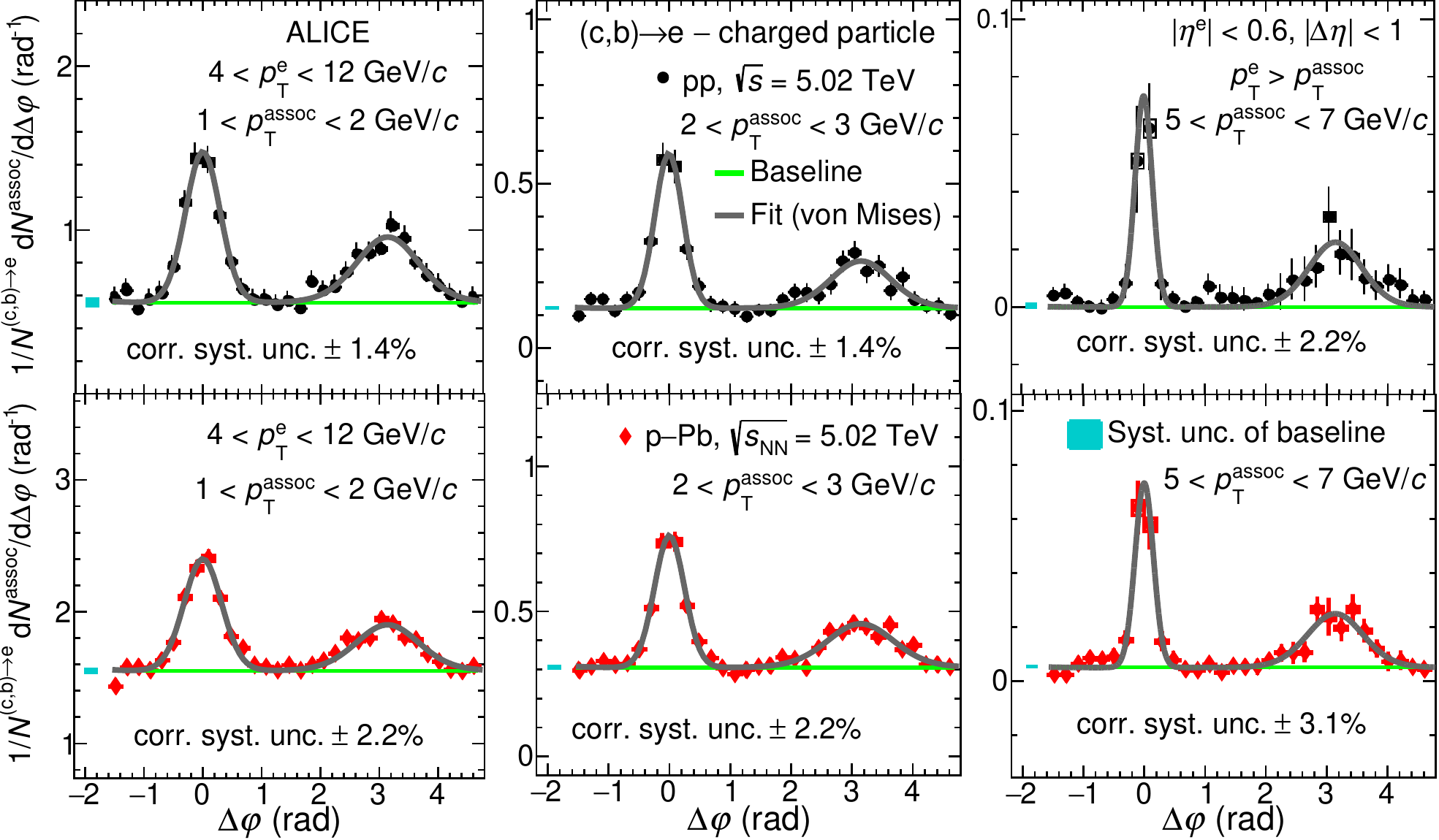}
}
\caption{The azimuthal-correlation distribution for $4 < \pt^{\rm e} < 12$ \GeVc ~fitted with a constant function for the baseline (green line) and von Mises functions for AS and NS peaks (grey curves) for different associated \pt ranges in pp collisions at $\sqrt{s}= 5.02$ TeV (top panels) and p--Pb collisions at $\sqrt{s_{\rm NN}}= 5.02$ TeV (bottom panels). The statistical (uncorrelated systematic) uncertainties are shown as vertical lines (empty boxes). The uncertainties of the baseline estimation are shown as solid boxes at $\dph \sim -2$ rad.}
\label{fig:DelphiWithPedpp}
\end{figure}

The azimuthal-correlation distributions for $|\Delta\eta|<1$ with trigger electron in the interval $4 < \pt^{\rm e} < 12$~\GeVc and for different associated particle \pt ranges together with their fit functions are shown in Fig.~\ref{fig:DelphiWithPedpp} (for selected $\pt^{\rm{assoc}}$ ranges) for pp (top panels) and p--Pb (bottom panels) collisions. The correlated systematic uncertainties, from the associated particle selection and mixed-event correction, are reported as text for each $\pt^{\rm{assoc}}$ interval. The baseline is shown by the horizontal green line. The absolute systematic uncertainty of the baseline estimation is shown as a solid box at $\dph \sim -2$~rad. The near- and away-side peaks are well described by the von Mises fit function in all $\pt^{\rm{assoc}}$ ranges. While the baseline contribution is higher in p--Pb collisions (due to the larger charged-particle multiplicity), its absolute value reduces with increasing $\pt^{\rm{assoc}}$ in both pp and p--Pb collisions. As a large fraction of the baseline is from the underlying event processes, the pairs contributing to it are dominated by low \pt particles.

\begin{figure}[!h]
\centering
\subfigure{
\includegraphics[scale=0.8]{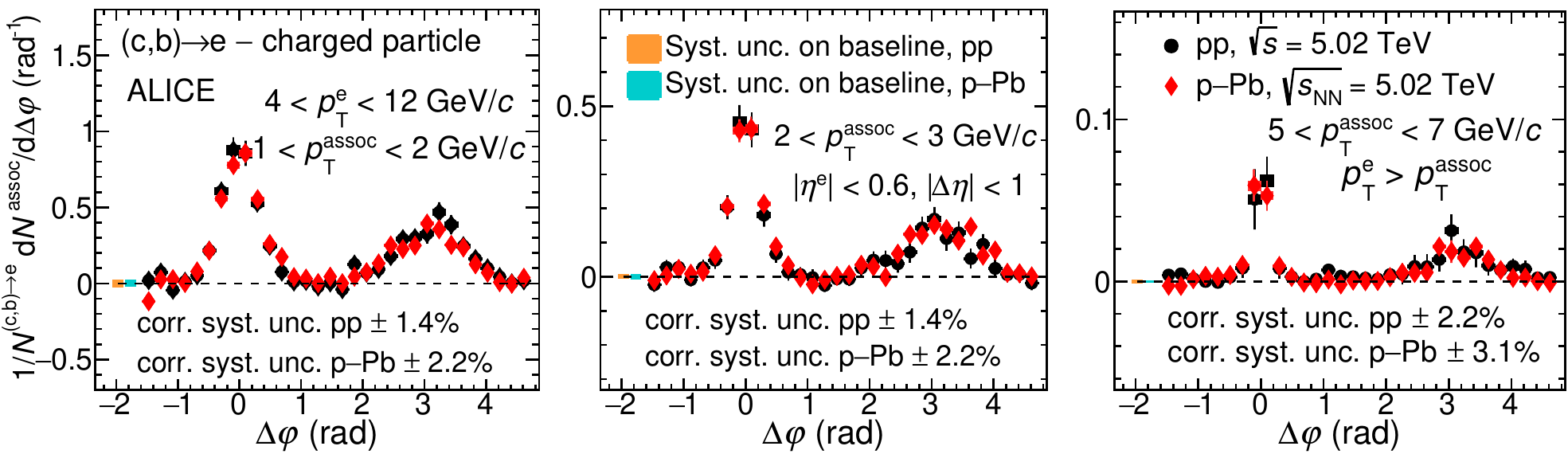}
}
\caption{Comparison of azimuthal-correlation distribution after baseline subtraction for $4 < \pt^{\rm e} < 12$ \GeVc ~and for different associated \pt ranges in pp collisions at $\sqrt{s} = 5.02$ TeV and p--Pb collisions at $\sqrt{s_{\rm{NN}}} = 5.02$ TeV. The statistical (uncorrelated systematic) uncertainties are shown as vertical lines (empty boxes). The uncertainties of the baseline estimation are shown as solid boxes at $\dph \sim -2$ rad.}
\label{fig:Delphi_pp_pPbrow}
\end{figure}

To compare the NS and AS peaks of the $\dph$ correlation distribution between pp and p--Pb collisions, the baseline-subtracted distributions from the two collision systems are shown together in Fig.~\ref{fig:Delphi_pp_pPbrow}, for $4 < \pt^{\rm e} < 12$ \GeVc ~and for different $\pt^{\rm{assoc}}$ ranges. It can be seen that the peak heights of the NS and AS decrease with increasing $\pt^{\rm{assoc}}$. A tendency for a more pronounced collimation of the NS peak with increasing $\pt^{\rm{assoc}}$ is visible. The profile of the correlation peaks is consistent in pp and p--Pb collisions within the statistical and systematic uncertainties.
This indicates that cold-nuclear matter effects do not impact heavy-quark fragmentation and hadronization in the measured \pt range, in minimum bias collisions. This observation is consistent with previous measurements of D-meson correlations with charged particles~\cite{ALICE:2019oyn, ALICE:2016clc}.

\begin{figure}[!h]
\centering
\subfigure{
\includegraphics[scale=0.7]{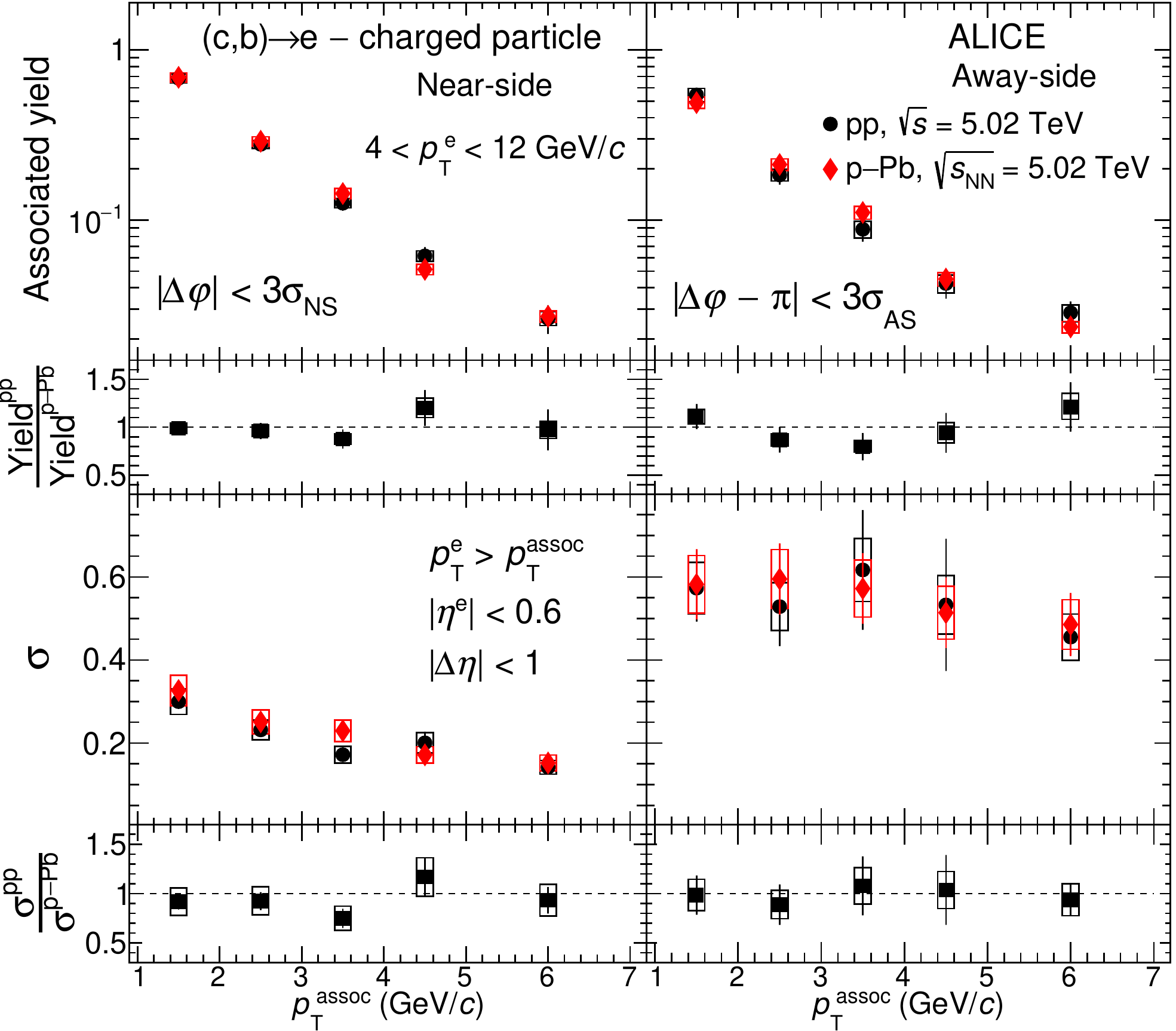}
}
\caption{Comparison of near- and away-side per-trigger yields (first row) and widths (third row) as a function of $\pt^{\rm{assoc}}$ for $4 < \pt^{\rm e} < 12$ \GeVc ~in pp collisions at $\sqrt{s} = 5.02$~TeV and p--Pb collisions at $\sqrt{s_{\rm{NN}}} = 5.02$~TeV. The ratios between pp and p--Pb yields and widths are shown in the second and fourth row, respectively. The statistical (systematic) uncertainties are shown as vertical lines (empty boxes).}
\label{fig:yieldsigma}
\end{figure}

To perform a quantitative comparison of the correlation peaks between pp and p--Pb collisions, the per-trigger NS and AS peak yields (first row) and widths (third row) are shown in Fig.~\ref{fig:yieldsigma}, superimposed for the two collision systems, as a function of $\pt^{\rm{assoc}}$ for $4 < \pt^{\rm e} < 12$ \GeVc. The ratios between pp and p--Pb yields (second row) and widths (fourth row) are also shown in this figure. The systematic uncertainties on the ratio of the yields and widths were obtained by considering all sources except for the baseline estimation as uncorrelated between pp and p--Pb collisions. The partially correlated uncertainty of the baseline estimation, obtained by using different fit functions, was estimated on the ratio. The total uncertainty was obtained by taking the quadratic sum of the correlated and uncorrelated uncertainties. While the NS and AS yields decrease with increasing $\pt^{\rm{assoc}}$ for both pp and p--Pb collisions, the measured yields are consistent within uncertainties between the two collision systems for all the $\pt^{\rm{assoc}}$ ranges, as can be seen in the ratio panels of Fig.~\ref{fig:yieldsigma}. The decrease in yields with increasing $\pt^{\rm{assoc}}$ can be understood considering that the heavy quarks have, on average, a hard fragmentation into heavy-flavor hadrons. As the remaining energy of heavy quarks is limited, it is far more likely that the associated particles accompanying the decay electron are preferentially produced at lower $\pt$.
The NS width values tend to decrease with increasing $\pt^{\rm{assoc}}$, with a value of about 0.3  at $\pt^{\rm{assoc}}=$ 1 \GeVc ~and narrowing to a value of roughly 0.15
at 6 \GeVc, with a significance of about $3\sigma$, for both pp and p--Pb collisions. The significance is calculated on the difference between the widths in the lowest and highest $\pt^{\rm{assoc}}$ intervals, taking into account both statistical and systematic uncertainties. The AS widths are independent of $\pt^{\rm{assoc}}$, and have a value of about 0.5. The NS peak distribution is closely connected to the fragmentation of the jet containing the trigger particle. The narrowing of the NS width with increasing $\pt^{\rm{assoc}}$ indicates that higher \pt particles tend to be closer to the jet-axis, whose direction can be approximated by the trigger electron. This is in turn related to higher 
$\pt$ emissions from the heavy quark being more collinear to it. The AS peak exhibits a lower sensitivity to the fragmentation of a specific heavy quark, as it can contain particles produced via the fragmentation of heavy quarks originating from processes, including next-to-leading order, that are not azimuthally back-to-back. These processes may have different relative fractions for different $\pt$ of heavy quarks. In the case of gluon splitting, the AS peak can also include particles originating from the recoil gluon, which are not directly associated with the heavy quarks produced in the event. Even in back-to-back processes, the correlation between the transverse momentum of the trigger electron and that of the opposite-side heavy quark, responsible for generating the AS peak through fragmentation, is significantly weaker than for the near-side peak. The NS and AS widths are similar in pp and p--Pb collisions, as can be seen in the ratio plots. 

\subsection{Comparison with predictions from MC event generators}

\begin{figure}[!h]
\centering
\subfigure{
\includegraphics[scale=0.8]{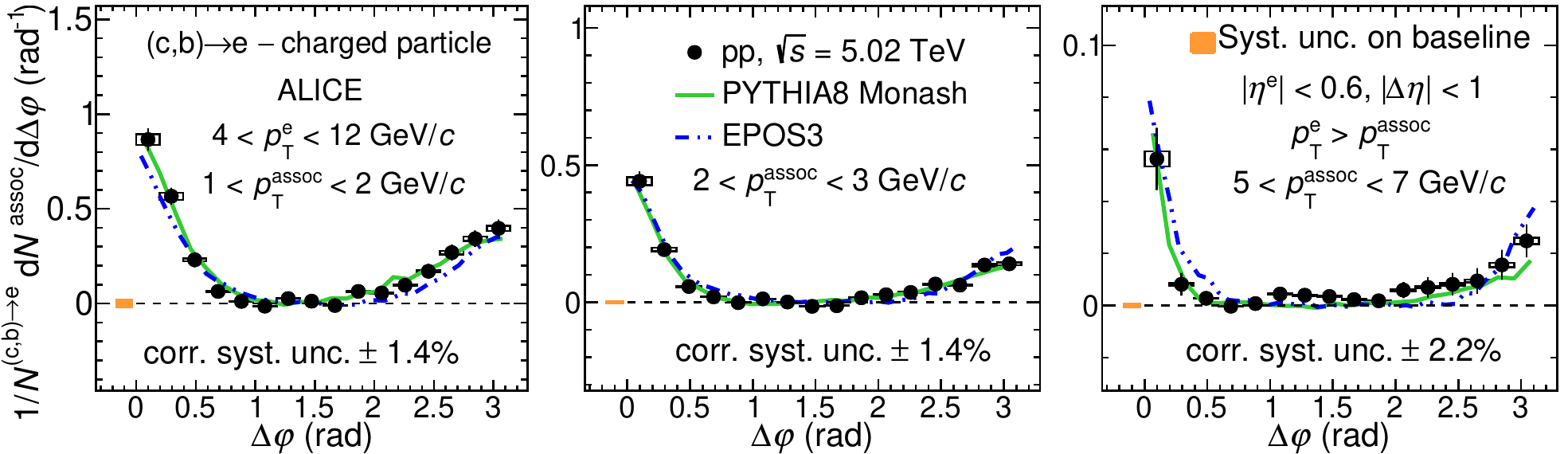}
}
\caption{Comparison of the azimuthal-correlation distribution with model predictions after baseline subtraction for $4 < \pt^{\rm e} < 12$ \GeVc ~in different $\pt^{\rm{assoc}}$ ranges in pp collisions at $\sqrt{s} = 5.02$~TeV. The statistical (uncorrelated systematic) uncertainties are shown as vertical lines (empty boxes). The uncertainties of the baseline are shown as solid boxes near $\dph \sim 0$ rad.}
\label{fig:Delphi_pp_models}
\end{figure}

\begin{figure}[!h]
\centering
\subfigure{
\includegraphics[scale=0.8]{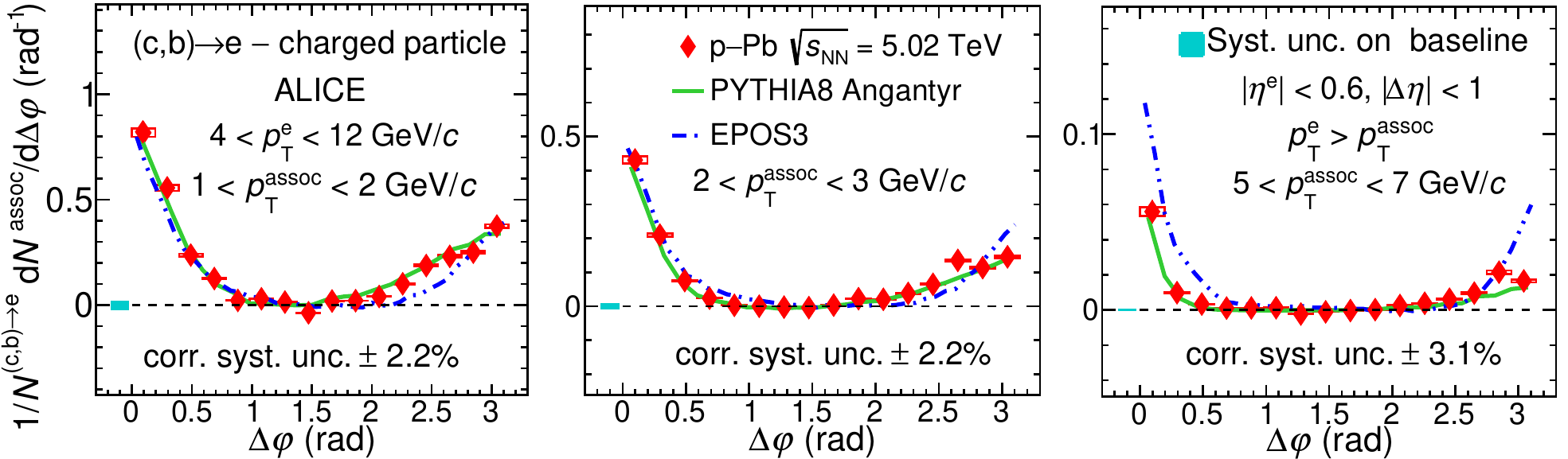}
}
\caption{Comparison of the azimuthal-correlation distribution with model predictions after baseline subtraction for $4 < \pt^{\rm e} < 12$ \GeVc ~in different $\pt^{\rm{assoc}}$ ranges in p--Pb collisions at $\sqrt{s_{\rm{NN}}} = 5.02$~TeV. The statistical (uncorrelated systematic) uncertainties are shown as vertical lines (empty boxes). The uncertainties of the baseline are shown as solid boxes near $\dph \sim 0$ rad.}
\label{fig:Delphi_pPb_models}
\end{figure}

\begin{figure}[!h]
\centering
\subfigure{
\includegraphics[scale=0.7]{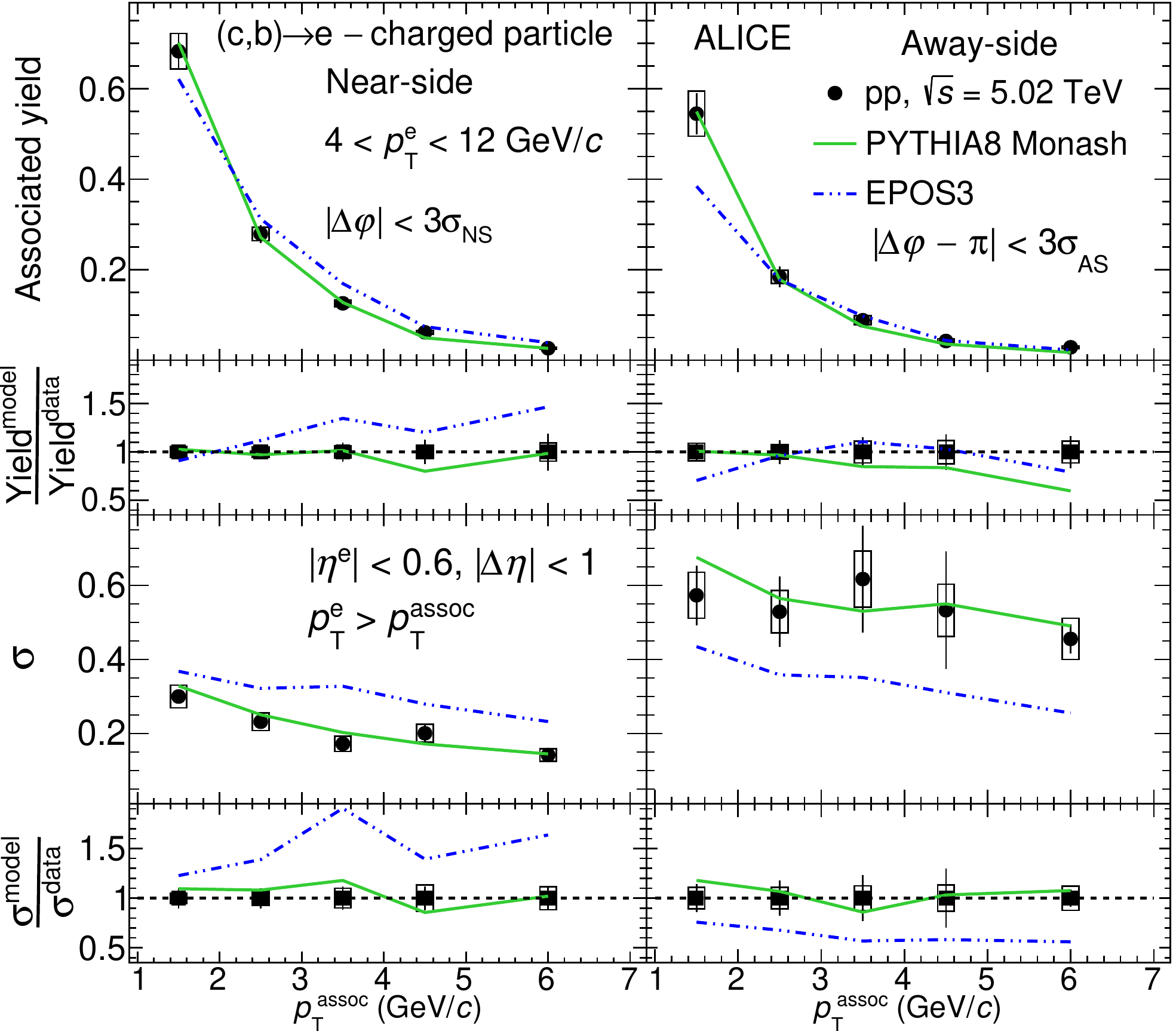}
}
\caption{Near- and away-side per-trigger yields (first row) and widths (third row) as a function of $\pt^{\rm{assoc}}$ for $4 < \pt^{\rm e} < 12$ \GeVc ~compared with predictions from PYTHIA8 Monash tune and EPOS3 in pp collisions at $\sqrt{s} = 5.02$~TeV. The ratios between model predictions and data are shown in the second and fourth row for the yields and widths, respectively. The statistical (systematic) uncertainties are shown as vertical lines (empty boxes).}
\label{fig:MCyieldsigma}
\end{figure}

\begin{figure}[h!]
\centering
\subfigure{
\includegraphics[scale=0.7]{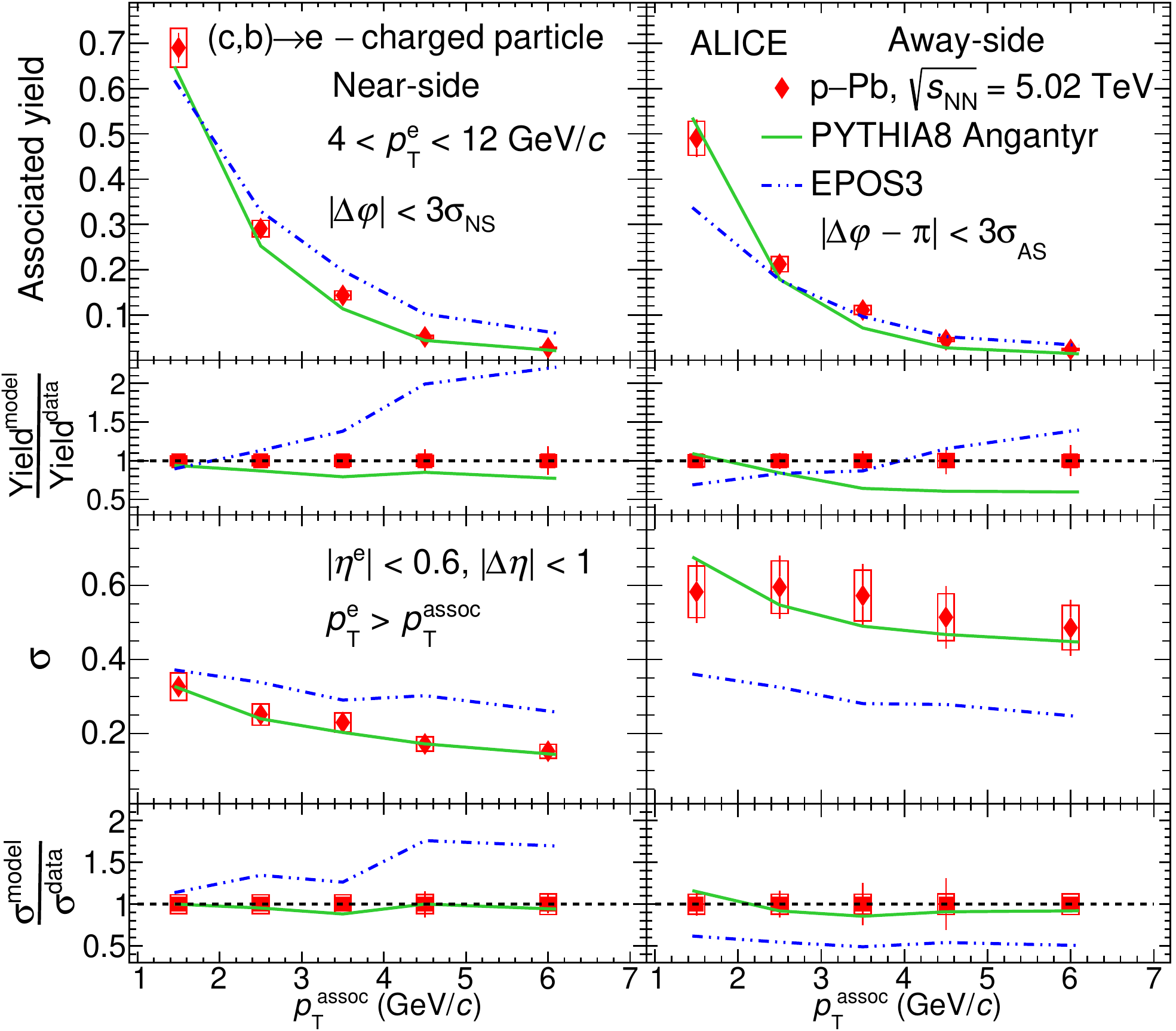}
}
\caption{Near- and away-side per-trigger yields (first row) and widths (third row) as a function of $\pt^{\rm{assoc}}$ for $4 < \pt^{\rm e} < 12$ \GeVc ~compared with predictions from PYTHIA8 Angantyr and EPOS3  in p--Pb collisions at $\sqrt{s_{\rm{NN}}} = 5.02$~TeV. The ratios between model predictions and data are shown in the second and fourth row for the yields and widths, respectively. The statistical (systematic) uncertainties are shown as vertical lines (empty boxes).}
\label{fig:pPbMCyieldsigma}
\end{figure}

The near- and away-side peaks of the azimuthal-correlation distribution in pp and p--Pb collisions are compared with predictions from different MC event generators. This allows verifying the implementation of the processes of charm- and beauty-quark production, fragmentation, and hadronization, which have an impact on the observables studied in this paper.
The models used to compare the measurement in pp collisions are PYTHIA8 with the Monash tune~\cite{Skands:2014pea,Sjostrand:2006za, Sjostrand:2007gs, Christiansen:2015yqa} and EPOS~3.117~\cite{Werner:2010aa,Drescher:2000ha}. The PYTHIA8 event generator is widely used in particle physics, as it provides an accurate description of high-energy collisions. It is capable of generating both hard and soft interactions, initial and final-state parton showers, particle fragmentation, and multi-partonic interactions. It also incorporates color reconnection mechanisms to rearrange color connections between quarks and gluons during hadronization. The prediction of these models for correlations of D mesons with charged particles can be found in Refs.~\cite{ALICE:2019oyn, ALICE:2016clc}. The p--Pb measurements are compared with PYTHIA8 Angantyr~\cite{Bierlich:2018xfw,Singh:2023rex}  and EPOS~3.117~\cite{Werner:2010aa,Drescher:2000ha} models.  The Angantyr~\cite{Bierlich:2018xfw,Singh:2023rex} model is used to simulate ultra-relativistic p--Pb collisions with the PYTHIA8 event generator.
As PYTHIA8 does not natively support collisions involving nuclei, this feature is implemented in the Angantyr model, which combines several nucleon--nucleon collisions to build a proton--nucleus (p--A) or nucleus--nucleus (A--A) collision. In this model, some modifications are made over the dynamics of pp collisions. The Angantyr model improves the inclusive definition of collision types of the FRITIOF model~\cite{Andersson:1986gw,Pi:1992ug}. In this model, a projectile nucleon can interact with several target nucleons where one primary collision looks like a typical pp non-diffractive (ND) collision. 
However, other target nucleons may also
undergo ND collisions with the projectile. 
The Angantyr model treats secondary ND collisions as modified single-diffractive (SD) interactions. For every p--A or A--A collision, nucleons are distributed randomly inside a nucleus according to a Glauber formalism similar to the one described in Ref.~\cite{Alvioli:2013vk}. This model is able to correctly reproduce final-state observables of heavy-ion collisions, i.e., multiplicity and \pt distributions~\cite{Singh:2021edu}. As collectivity is not incorporated in this model, its predictions serve as a baseline for studying observables sensitive to collective behavior in p--A and A--A systems. 
For PYTHIA8 simulations, the correlation distributions for electrons from charm- and beauty-hadron decays are obtained separately, 
and summed after weighting their relative fractions based on FONLL calculations~\cite{Cacciari:1998it,Cacciari:2001td,Cacciari:2012ny,ALICE:2014aev}.

The EPOS3 event generator is largely used for the description of ultra-relativistic heavy-ion collisions. It employs a core-corona description of the fireball produced in these collisions: in the ``core", its inner part, a quark--gluon plasma is formed, which follows a hydrodynamic behavior, while in the external regions of the ``corona" the partons fragment and hadronize independently.
A study of radial flow performed with the EPOS3  event generator in proton--proton collisions at $\sqrt{s}$ = 7 TeV~\cite{Ortiz:2016kpz} has shown that the energy density reached in such collisions is large enough to grant the applicability of the hydrodynamic evolution to the core of the collision.

In the models, the azimuthal correlation function of trigger electrons from charm- and beauty-hadron decays with charged particles is evaluated using the same prescriptions applied for data analysis in terms of kinematic and particle-species selections. The peak properties of the correlation functions are obtained by following the same approach employed in data, i.e., by fitting the distributions with two von Mises functions and a constant term.

In Figs.~\ref{fig:Delphi_pp_models} and~\ref{fig:Delphi_pPb_models}, the baseline-subtracted azimuthal-correlation distribution measured in pp and p--Pb collisions, reflected in the $0 < \Delta\varphi < \pi $ range, is compared with predictions from PYTHIA8 and EPOS3 generators for $4 < \pt^{\rm e} < 12$ \GeVc in three different $\pt^{\rm{assoc}}$ ranges. The comparison for the remaining $\pt^{\rm{assoc}}$ ranges is shown in Appendix~\ref{SuppleFigsAddpTass}.
From this qualitative comparison, both MC generators give a good overall description of the data in all the $\pt^{\rm{assoc}}$ intervals, even though the EPOS3 predictions show some deviation from the measured NS and AS peaks in the highest $\pt^{\rm{assoc}}$ interval. The peak yields and widths extracted from the measured distribution are also compared with model predictions in Figs.~\ref{fig:MCyieldsigma} and~\ref{fig:pPbMCyieldsigma} for pp and p--Pb collisions, respectively. From here on, PYTHIA8/Angantyr will be used to refer to PYTHIA8 Monash simulations in pp collisions and PYTHIA8 Angantyr simulations in p--Pb collisions together. PYTHIA8/Angantyr simulations provide NS and AS yields decreasing with increasing $\pt^{\rm{assoc}}$ and are consistent with the data within statistical and systematic uncertainties. The NS widths simulated using PYTHIA8/Angantyr decrease with increasing $\pt^{\rm{assoc}}$, which are consistent with the data in both collision systems. The AS widths show a slightly decreasing trend with $\pt^{\rm{assoc}}$ that is consistent with data within statistical and systematic uncertainties in both collision systems.  
The NS and AS yields predicted by the EPOS3 model qualitatively describe the data within statistical and systematic uncertainties in pp collisions. In p--Pb collisions, the NS yield is overestimated at high $\pt^{\rm{assoc}}$ while the AS yield is consistent with data within statistical and systematic uncertainties.
The EPOS3 simulations overestimate the NS widths and underestimate the AS widths for all $\pt^{\rm{assoc}}$ ranges in pp and p--Pb collisions.  

\subsection{Dependence of the correlation distribution on the $\pt^{\rm{e}}$
} \label{sec:pTeBinSplitStudies}

\begin{figure}[!h]
\centering
\includegraphics[scale=0.7]{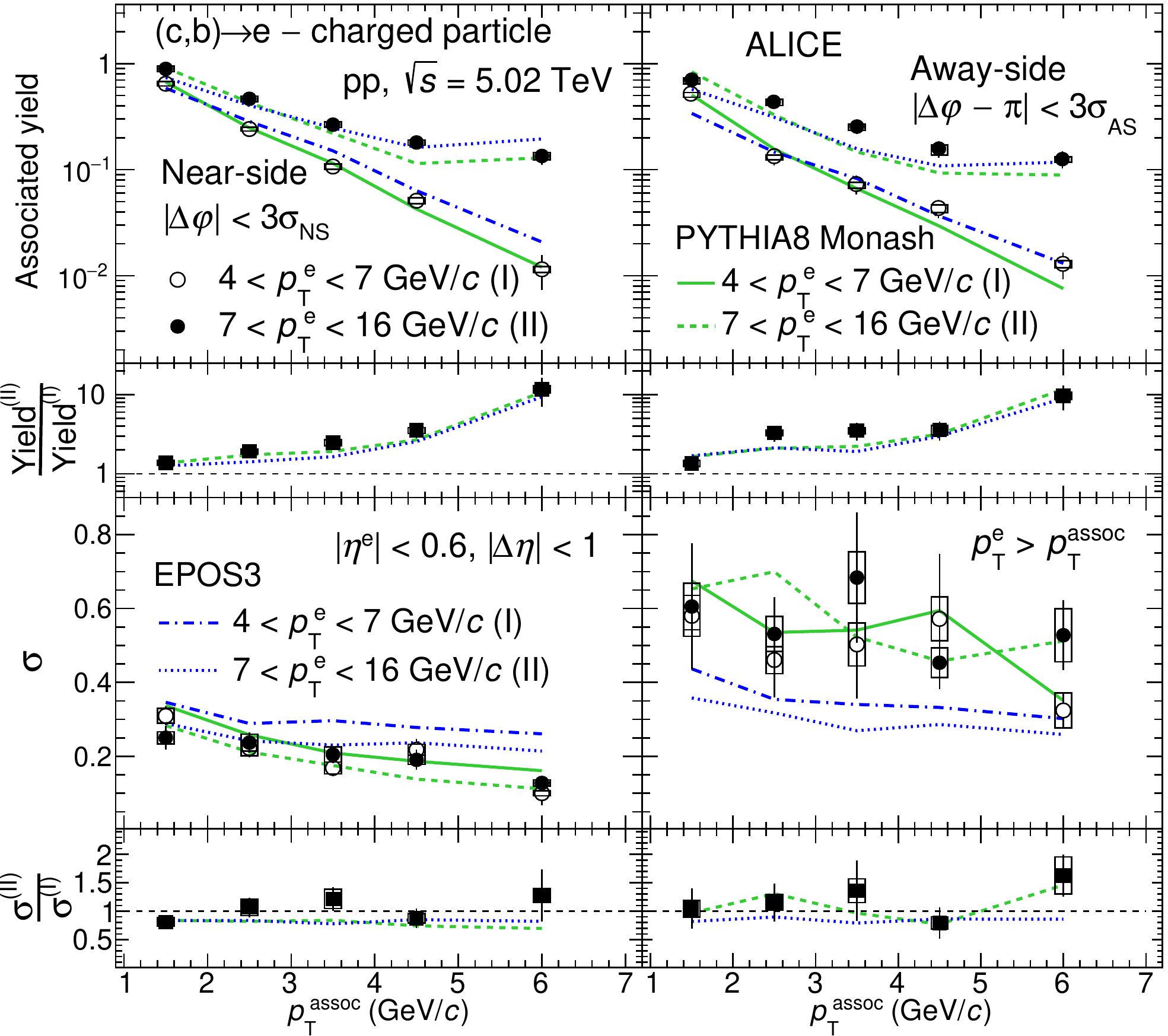}
\caption{Comparison of NS and AS per-trigger yields (first row) and widths (third row) for two $\pt^{\rm{e}}$ ranges $4 < \pt^{\rm e} < 7$ \GeVc ~and $7 < \pt^{\rm e} < 16$ \GeVc, as a function of $\pt^{\rm{assoc}}$ in pp collisions. The ratios between the $7 < \pt^{\rm e} < 16$ \GeVc and $4 < \pt^{\rm e} < 7$ \GeVc yields and widths are shown in the second and fourth rows, respectively. The data are compared with PYTHIA8 Monash and EPOS3 predictions. The statistical (systematic) uncertainties are shown as vertical lines (empty boxes).}
\label{fig:yieldsigmapp47_716}
\end{figure}

\begin{figure}[!h]
\centering
\includegraphics[scale=0.7]{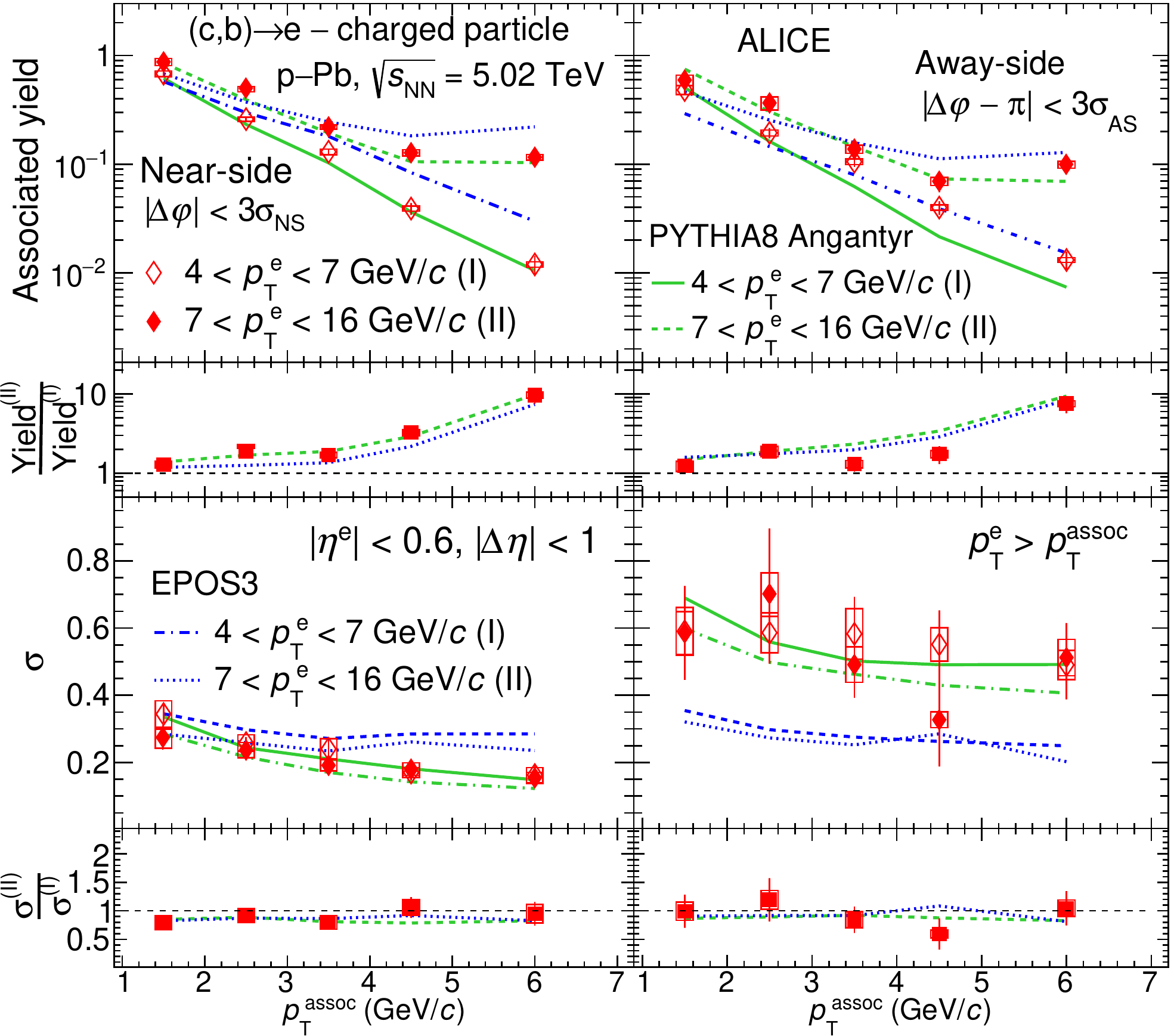}

\caption{Comparison of NS and AS per-trigger yields (first row) and widths (third row) for two $\pt^{\rm{e}}$ ranges $4 < \pt^{\rm e} < 7$ \GeVc ~and $7 < \pt^{\rm e} < 16$ \GeVc, as a function of $\pt^{\rm{assoc}}$ in p--Pb collisions. The ratios between the $7 < \pt^{\rm e} < 16$ \GeVc and $4 < \pt^{\rm e} < 7$ \GeVc yields and widths are shown in the second and fourth rows, respectively. The data are compared with PYTHIA8 Angantyr and EPOS3 predictions. The statistical (systematic) uncertainties are shown as vertical lines (empty boxes).}
\label{fig:yieldsigmapPb47_716}
\end{figure}

\begin{figure}[!h]
\centering
\includegraphics[scale=0.7]{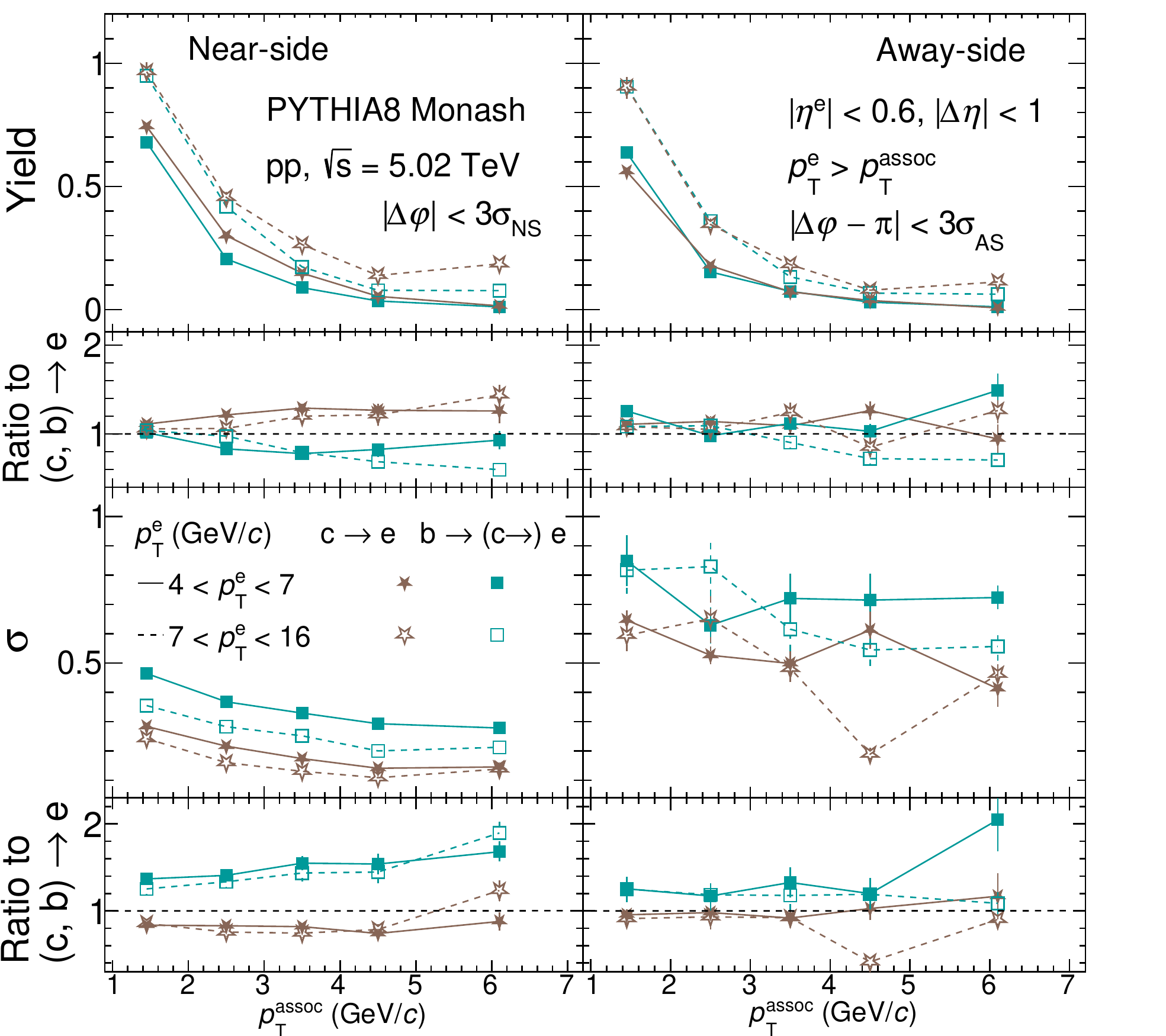}

\caption{Comparison of PYTHIA8 Monash prediction for NS and AS per-trigger yields (first row) and widths (third row) in the two $\pt^{\rm{e}}$ ranges $4 < \pt^{\rm e} < 7$ \GeVc ~and $7 < \pt^{\rm e} < 16$ \GeVc for electrons from charm- and beauty-hadron decays, as a function of $\pt^{\rm{assoc}}$ in pp collisions. The ratios to c, b $\rightarrow$ e yields and widths are shown in the second and fourth rows, respectively. The statistical uncertainties are shown as vertical lines.}
\label{fig:pythiaCB}
\end{figure}

The relative fractions of electrons produced by charm- and beauty-hadron decays have a strong \pt dependence~\cite{ALICE:2014aev}. The fraction of electrons from beauty-hadron decays at $\pt^{\rm e}$ = 4 \GeVc ~accounts for about 40\% of the HFe yield, increasing to 60--70\% for $\pt^{\rm e} >8$ \GeVc. A dependence of the correlation distribution on the flavor of the quark from which the trigger electron originates can be expected, due to the different fragmentation of charm and beauty quarks and different fraction of LO and NLO processes involved in their production. The correlation distributions for electrons from a given quark flavor can also have a trigger-particle \pt ~dependence due to the different energy of the original parton, and different relative contribution of LO and NLO production processes for the hard scattering producing the parton. These effects are studied by measuring the correlation distributions for trigger electrons in the \pt ranges $4 < \pt^{\rm e} < 7$ \GeVc ~and $7 < \pt^{\rm e} < 16$ \GeVc, where the latter $\pt^{\rm e}$ range is dominated by electrons from beauty-hadron decays. The azimuthal correlation distributions for these two $\pt^{\rm e}$ ranges are presented in Appendix~\ref{SuppleFigs}. The NS and AS yields and widths for the two $\pt^{\rm e}$ intervals are obtained following the same procedure described in Sec. 3.  

The comparisons of the yields (first row) and widths (third row) for the two $\pt^{\rm e}$ bins are shown in Figs.~\ref{fig:yieldsigmapp47_716} and~\ref{fig:yieldsigmapPb47_716} for pp and p--Pb collisions, respectively. The per-trigger NS and AS yields are systematically higher for the $7 < \pt^{\rm e} < 16$ \GeVc ~range compared to the values obtained for $4 < \pt^{\rm e} < 7$ \GeVc, for both pp and p--Pb collisions. The ratio between the $7 < \pt^{\rm e} < 16$ \GeVc and $4 < \pt^{\rm e} < 7$ \GeVc yields is shown in the second row of Figs.~\ref{fig:yieldsigmapp47_716} and~\ref{fig:yieldsigmapPb47_716}. It can be observed that the yield is higher for the higher $\pt^{\rm{e}}$ interval, and the ratio increases from 1.3 at low $\pt^{\rm{assoc}}$ to $\sim 10$ in the highest $\pt^{\rm{assoc}}$ interval, for both pp and p--Pb collisions. This can be explained by considering that higher-\pt electrons are typically produced by more energetic heavy quarks, and the additional parton energy on average leads to a larger number of associated fragmentation particles. 

While the NS width values decrease with $\pt^{\rm{assoc}}$, they are similar for the two trigger electron $\pt$ ranges. The AS widths are also observed to be similar for the two trigger electron $\pt$ ranges and to have an almost flat trend with $\pt^{\rm{assoc}}$. It should be noted that the kinematic bias induced due to the condition of $\pt^{\rm{assoc}} < \pt^{\rm e}$ affects the correlation distributions for the two trigger electron $\pt$ ranges differently. While none of the correlation distributions for higher $\pt^{\rm{e}}$ interval are affected by the bias, the distributions for $4 < \pt^{\rm e} < 7$ \GeVc and $4 < \pt^{\rm{assoc}} < 7$ \GeVc would miss some associated particles because of the selection condition.

The NS and AS yields and widths of the correlation distributions as a function of $\pt^{\rm{assoc}}$ for the two $\pt^{\rm{e}}$~ranges are compared with PYTHIA8/Angantyr and EPOS3 MC simulations for pp and p--Pb collisions. The PYTHIA8/Angantyr predictions describe the data within uncertainties for both $\pt^{\rm{e}}$ ranges. The NS and AS yields from EPOS3 are consistent with data for both $\pt^{\rm{e}}$ intervals. The trend of NS width from EPOS3 is slightly flatter as a function of $\pt^{\rm{assoc}}$ compared to that of the data. Similar to what was observed for $4 < \pt^{\rm e} < 12$ \GeVc, the NS width is overestimated, while the AS width is underestimated compared to data for both $\pt^{\rm e}$ ranges. The ratio of the yields and widths of the two $\pt^{\rm e}$ ranges are well described by both MC event generators. 

To understand the effect of the different charm and beauty fragmentation on the observed $\pt^{\rm{e}}$ ~dependence, the correlation distributions were obtained for electrons from charm- and beauty-hadron decays separately for the two $\pt^{\rm e}$ intervals using PYTHIA8 MC simulations. The NS and AS yields and widths of the correlation distributions for electrons from charm- and beauty-hadron decays, and their ratios to the combined ones (HFe), are shown in Fig.~\ref{fig:pythiaCB}. For both $\pt^{\rm e}$ intervals, the NS yields for trigger electrons from beauty-hadron decays are lower than those from charm-hadron decays, by about 5\% for the first $\pt^{\rm{assoc}}$ interval, with a tendency for an increased difference for larger $\pt^{\rm{assoc}}$, about 40\% for the last $\pt^{\rm{assoc}}$ range. This can be expected due to the harder fragmentation of beauty quarks to beauty hadrons compared to that of charm quarks, with less energy remaining for the production of other particles in the parton shower. This indicates that the yield increase at higher $\pt^{\rm{e}}$  observed in Figs.~\ref{fig:yieldsigmapp47_716} and~\ref{fig:yieldsigmapPb47_716} is largely due to the higher energy of the initial heavy quark.  
The NS and AS widths of the correlation distributions decrease with increasing $\pt^{\rm e}$ for both charm- and beauty-hadron decays, but the widths for electrons from beauty-hadron decays are wider than for electrons from charm-hadron decays for both $\pt^{\rm e}$ intervals. These two opposing effects lead to similar widths for the two $\pt^{\rm{e}}$ intervals in Figs.~\ref{fig:yieldsigmapp47_716} and~\ref{fig:yieldsigmapPb47_716}.


\section{Summary}
\label{sec:Summary}

Measurements of azimuthal-correlation functions of heavy-flavor hadron decay electrons with charged particles in pp and p--Pb collisions at $\sqrt{s_{\rm{NN}}} = 5.02$ TeV have been reported. The correlation distributions were obtained for trigger electrons in the range $4 < \pt^{\rm e} < 12$ \GeVc, and for different associated particle \pt ranges between 1 and 7 \GeVc. 
The azimuthal distributions were fitted with a constant and two von Mises functions in order to characterize the near- and away-side peaks. 

The evolution of the near- and away-side peaks of the correlation functions in pp and p--Pb collisions is found to be similar in all the considered kinematic ranges. This suggests that the modification of the fragmentation and hadronization of heavy quarks due to cold-nuclear-matter effects is indistinguishable within the current precision of the measurements. The extracted near- and away-side per-trigger yields and widths in pp and p--Pb collisions are presented as a function of associated particle \pt, which provide access to the momentum distributions of the particles produced in the fragmentation of the hard parton, and allow for a differential study of the jet angular profile. The per-trigger yields decrease with increasing $\pt^{\rm{assoc}}$ and are consistent between pp and p--Pb collisions. While the near-side width tends to decrease with increasing $\pt^{\rm{assoc}}$, the away-side width does not show a pronounced trend with $\pt^{\rm{assoc}}$ for both collision systems. 
The $\dph$ distributions, per-trigger yields, and widths in pp and p--Pb collisions are compared with predictions from PYTHIA8 (with Monash tune for pp and using the Angantyr model for p--Pb collisions), and EPOS3  Monte Carlo event generators. The PYTHIA8 predictions provide the best description of the data for both yields and widths of the near- and away-side peaks. For the current implementation of the EPOS3 model, the yields are similar to those obtained from data, while the near- and away-side widths are overestimated and underestimated, respectively. 

The relative fractions of electrons from charm- and beauty-hadron decays have a strong \pt ~dependence. This feature was exploited by studying the correlation distribution for the kinematic regions, $4 < \pt^{\rm e} < 7$ \GeVc ~and $7 < \pt^{\rm e} < 16$ \GeVc, where the latter $\pt^{\rm{e}}$~range is dominated by beauty-hadron decays.
 
For both collision systems studied, the per-trigger yields are systematically larger for the $7 < \pt^{\rm e} < 16$ \GeVc range compared to the $4 < \pt^{\rm e} < 7$ interval due to the larger energy of the initial heavy quark, which allows for the production of more particles in the parton shower. This effect dominates over the increased beauty-origin contribution of the trigger electrons in the $7 < \pt^{\rm e} < 16$ \GeVc range, which according to PYTHIA8 studies are characterized by lower correlation peak yields than those of electrons originating from charm. The near- and away-side widths are observed to be similar for both trigger electron \pt ranges, for pp and p--Pb collisions.  
PYTHIA8 studies indicates that this is due to competing effects, where the larger boost of the initial heavy quark leads to a stronger collimation of the peaks with increasing $\pt^{\rm e}$ for both charm- and beauty-origin contributions, compensating the broader peak widths for trigger electrons originating from beauty-hadron decays, whose contribution increases with $\pt^{\rm e}$.

The reported results constitute a reference for future measurements in Pb--Pb collisions at the same center-of-mass energy. The study of the modifications of the correlation functions in Pb--Pb collisions in the presence of QGP can provide a deeper understanding of heavy-quark dynamics inside the hot QCD medium~\cite{ALICE:2022wwr}. 



\newenvironment{acknowledgement}{\relax}{\relax}
\begin{acknowledgement}
\section*{Acknowledgements}

The ALICE Collaboration would like to thank all its engineers and technicians for their invaluable contributions to the construction of the experiment and the CERN accelerator teams for the outstanding performance of the LHC complex.
The ALICE Collaboration gratefully acknowledges the resources and support provided by all Grid centres and the Worldwide LHC Computing Grid (WLCG) collaboration.
The ALICE Collaboration acknowledges the following funding agencies for their support in building and running the ALICE detector:
A. I. Alikhanyan National Science Laboratory (Yerevan Physics Institute) Foundation (ANSL), State Committee of Science and World Federation of Scientists (WFS), Armenia;
Austrian Academy of Sciences, Austrian Science Fund (FWF): [M 2467-N36] and Nationalstiftung f\"{u}r Forschung, Technologie und Entwicklung, Austria;
Ministry of Communications and High Technologies, National Nuclear Research Center, Azerbaijan;
Conselho Nacional de Desenvolvimento Cient\'{\i}fico e Tecnol\'{o}gico (CNPq), Financiadora de Estudos e Projetos (Finep), Funda\c{c}\~{a}o de Amparo \`{a} Pesquisa do Estado de S\~{a}o Paulo (FAPESP) and Universidade Federal do Rio Grande do Sul (UFRGS), Brazil;
Bulgarian Ministry of Education and Science, within the National Roadmap for Research Infrastructures 2020-2027 (object CERN), Bulgaria;
Ministry of Education of China (MOEC) , Ministry of Science \& Technology of China (MSTC) and National Natural Science Foundation of China (NSFC), China;
Ministry of Science and Education and Croatian Science Foundation, Croatia;
Centro de Aplicaciones Tecnol\'{o}gicas y Desarrollo Nuclear (CEADEN), Cubaenerg\'{\i}a, Cuba;
Ministry of Education, Youth and Sports of the Czech Republic, Czech Republic;
The Danish Council for Independent Research | Natural Sciences, the VILLUM FONDEN and Danish National Research Foundation (DNRF), Denmark;
Helsinki Institute of Physics (HIP), Finland;
Commissariat \`{a} l'Energie Atomique (CEA) and Institut National de Physique Nucl\'{e}aire et de Physique des Particules (IN2P3) and Centre National de la Recherche Scientifique (CNRS), France;
Bundesministerium f\"{u}r Bildung und Forschung (BMBF) and GSI Helmholtzzentrum f\"{u}r Schwerionenforschung GmbH, Germany;
General Secretariat for Research and Technology, Ministry of Education, Research and Religions, Greece;
National Research, Development and Innovation Office, Hungary;
Department of Atomic Energy Government of India (DAE), Department of Science and Technology, Government of India (DST), University Grants Commission, Government of India (UGC) and Council of Scientific and Industrial Research (CSIR), India;
National Research and Innovation Agency - BRIN, Indonesia;
Istituto Nazionale di Fisica Nucleare (INFN), Italy;
Japanese Ministry of Education, Culture, Sports, Science and Technology (MEXT) and Japan Society for the Promotion of Science (JSPS) KAKENHI, Japan;
Consejo Nacional de Ciencia (CONACYT) y Tecnolog\'{i}a, through Fondo de Cooperaci\'{o}n Internacional en Ciencia y Tecnolog\'{i}a (FONCICYT) and Direcci\'{o}n General de Asuntos del Personal Academico (DGAPA), Mexico;
Nederlandse Organisatie voor Wetenschappelijk Onderzoek (NWO), Netherlands;
The Research Council of Norway, Norway;
Commission on Science and Technology for Sustainable Development in the South (COMSATS), Pakistan;
Pontificia Universidad Cat\'{o}lica del Per\'{u}, Peru;
Ministry of Education and Science, National Science Centre and WUT ID-UB, Poland;
Korea Institute of Science and Technology Information and National Research Foundation of Korea (NRF), Republic of Korea;
Ministry of Education and Scientific Research, Institute of Atomic Physics, Ministry of Research and Innovation and Institute of Atomic Physics and University Politehnica of Bucharest, Romania;
Ministry of Education, Science, Research and Sport of the Slovak Republic, Slovakia;
National Research Foundation of South Africa, South Africa;
Swedish Research Council (VR) and Knut \& Alice Wallenberg Foundation (KAW), Sweden;
European Organization for Nuclear Research, Switzerland;
Suranaree University of Technology (SUT), National Science and Technology Development Agency (NSTDA), Thailand Science Research and Innovation (TSRI) and National Science, Research and Innovation Fund (NSRF), Thailand;
Turkish Energy, Nuclear and Mineral Research Agency (TENMAK), Turkey;
National Academy of  Sciences of Ukraine, Ukraine;
Science and Technology Facilities Council (STFC), United Kingdom;
National Science Foundation of the United States of America (NSF) and United States Department of Energy, Office of Nuclear Physics (DOE NP), United States of America.
In addition, individual groups or members have received support from:
European Research Council, Strong 2020 - Horizon 2020, Marie Sk\l{}odowska Curie (grant nos. 950692, 824093, 896850), European Union;
Academy of Finland (Center of Excellence in Quark Matter) (grant nos. 346327, 346328), Finland;
Programa de Apoyos para la Superaci\'{o}n del Personal Acad\'{e}mico, UNAM, Mexico.

\end{acknowledgement}

\bibliographystyle{utphys}   
\bibliography{bibliography.bib,alice_papers.bib}

\newpage
\appendix
\section{Supplemental Figures} \label{SuppleFigs}
\begin{figure}[!h]
\centering
\subfigure{
\includegraphics[scale=0.4]{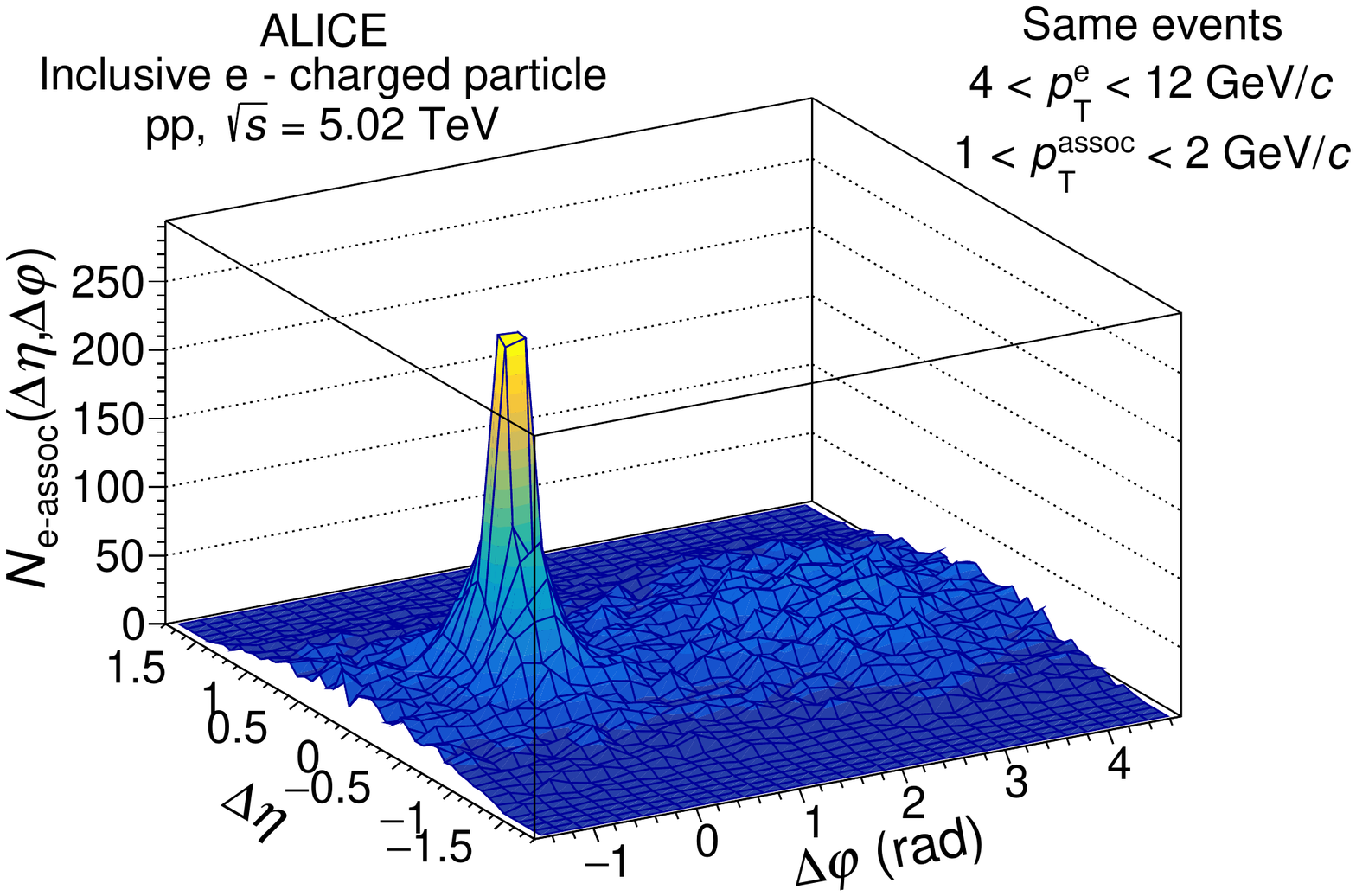}
\includegraphics[scale=0.4]{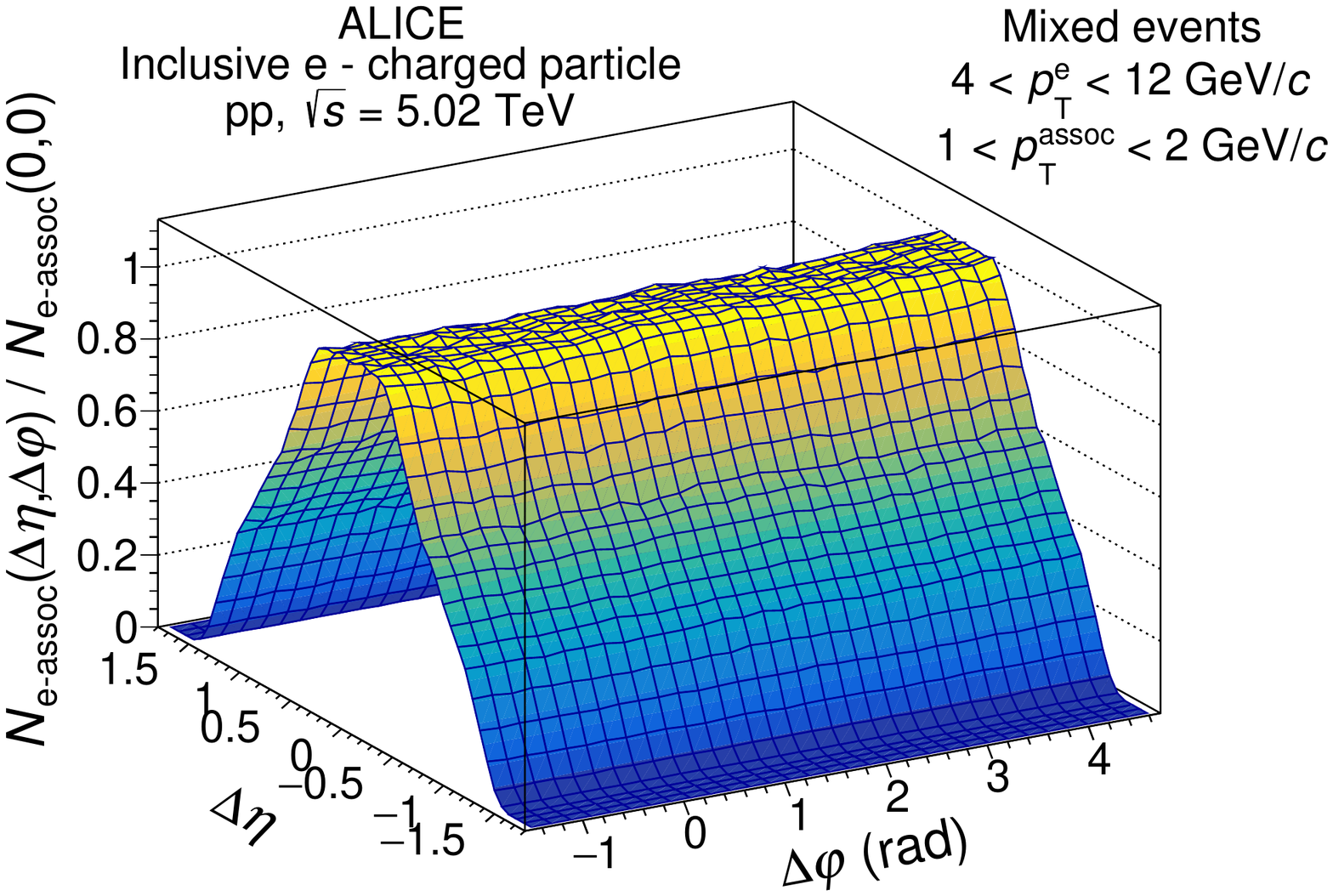}
}
\subfigure{
\includegraphics[scale=0.4]{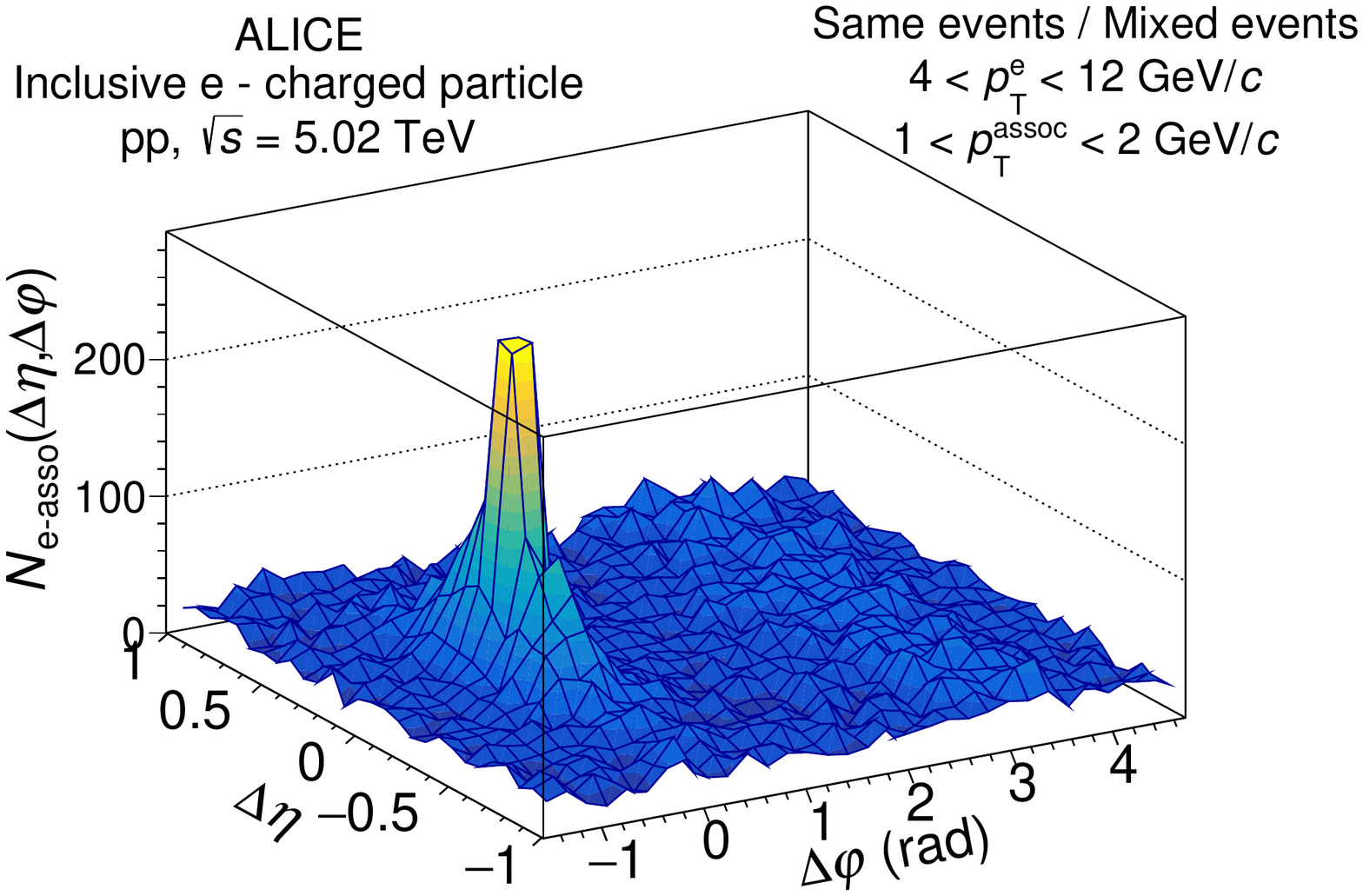}
}
\caption{Example of measured same-event (top-left), mixed-event (top-right), and corrected (bottom) correlation distribution, for $4 < p_{\rm T}^{\rm e} < 12$ \GeVc ~and $1 < p_{\rm T}^{\rm assoc} < 2$ \GeVc in pp collisions at $\sqrt{s} =$ 5.02 TeV.}
\label{fig:MEcorrectionpp}
\end{figure}

\begin{figure}[!h]
\centering
\subfigure{
\includegraphics[scale=0.8]{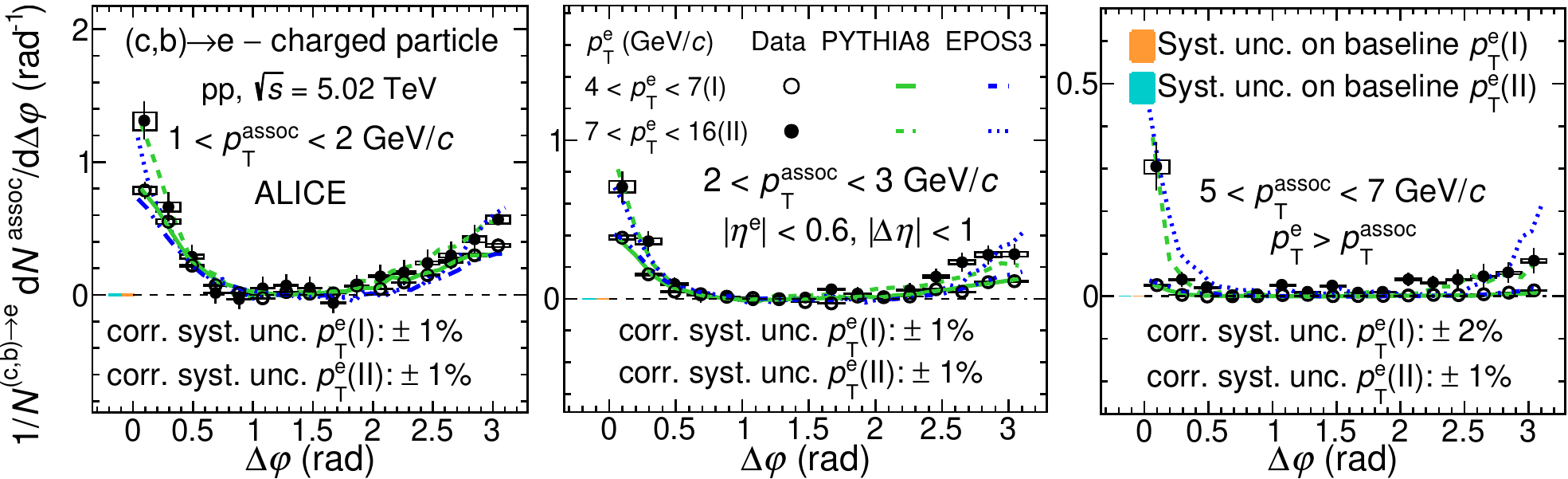}
}
\caption{Azimuthal-correlation distributions after baseline subtraction for two $\pt^{\rm e}$ intervals, $4 < \pt^{\rm e} < 7$ \GeVc and $7 < \pt^{\rm e} < 16$ \GeVc, and for different associated \pt ranges within $1 < \pt^{\rm{assoc}} < 7$ \GeVc compared with predictions from PYTHIA8 Monash and EPOS3 in pp collisions at $\sqrt{s} =$ 5.02 TeV. The statistical (systematic) uncertainties are shown as vertical lines (empty boxes). The uncertainties of the baseline are shown as solid boxes at $\dph \sim 0$ rad.}
\label{fig:Delphi_pppT47_716}
\end{figure}

\begin{figure}[!h]
\centering
\subfigure{
\includegraphics[scale=0.8]{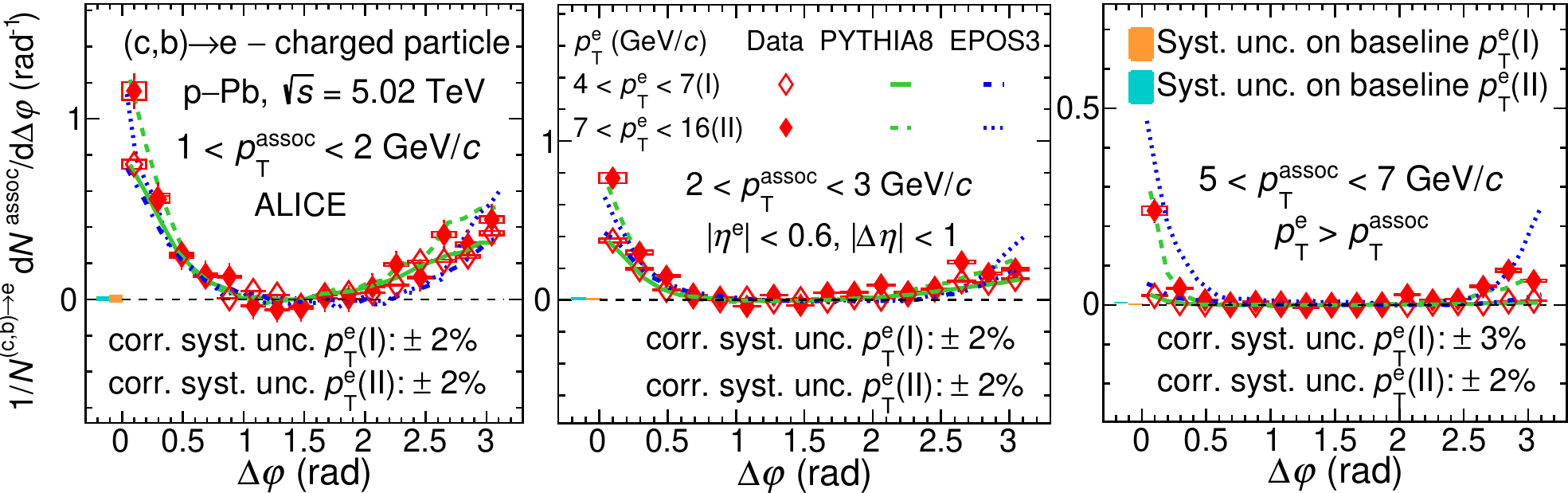}
}
\caption{Azimuthal-correlation distributions after baseline subtraction for two $\pt^{\rm e}$ intervals, $4 < \pt^{\rm e} < 7$ \GeVc and $7 < \pt^{\rm e} < 16$ \GeVc, and for different associated \pt ranges within $1 < \pt^{\rm{assoc}} < 7$ \GeVc compared with predictions from PYTHIA8 Angantyr and EPOS3 in p--Pb collisions at $\sqrt{s_{\rm{NN}}} =$ 5.02 TeV. The statistical (systematic) uncertainties are shown as vertical lines (empty boxes). The uncertainties of the baseline are shown as solid boxes at $\dph \sim 0$ rad.}
\label{fig:Delphi_pPpT47_716}
\end{figure}

In this appendix, some supplemental figures are reported. In particular, Fig.~\ref{fig:MEcorrectionpp} illustrates some details about the analysis steps described in Sec.~\ref{sec:MEcorrection}, while Figs.~\ref{fig:Delphi_pppT47_716} and~\ref{fig:Delphi_pPpT47_716} support the discussion reported in Sec.~\ref{sec:pTeBinSplitStudies} about the comparison of the measured correlation distributions with MC event generators.

\newpage

\section{Supplemental figures with additional $\pt^{\rm{assoc}}$ ranges}\label{SuppleFigsAddpTass}

\begin{figure}[!h]
\centering
\subfigure{
\includegraphics[scale=0.8]{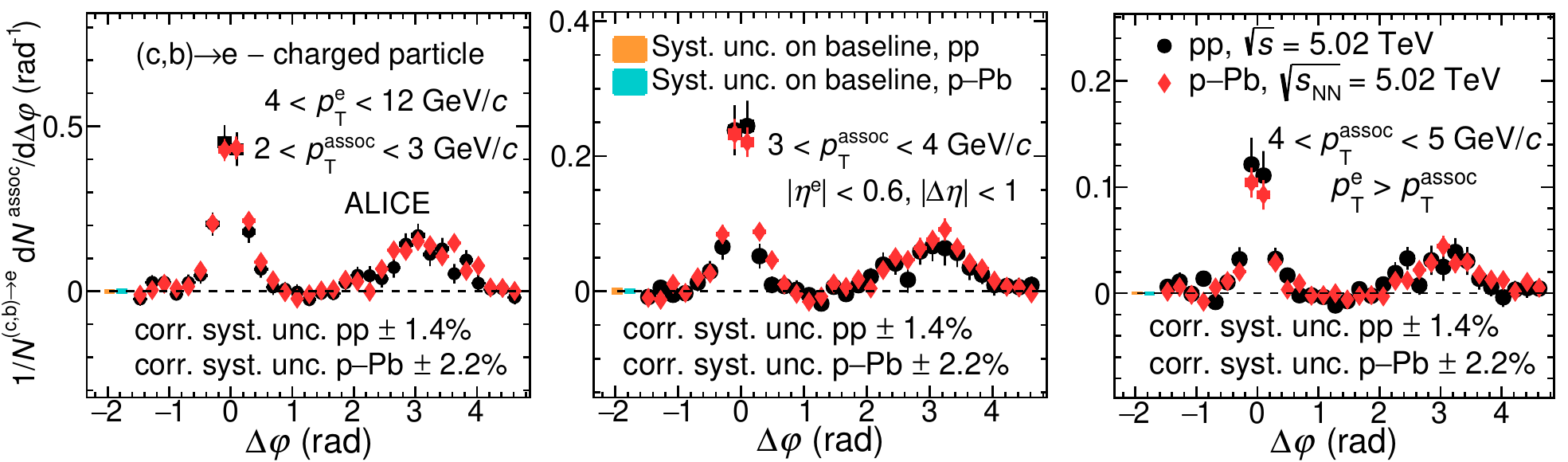}
}
\caption{Azimuthal-correlation distributions after baseline subtraction for $4 < \pt^{\rm e} < 12$ \GeVc and for different associated \pt ranges in pp collisions at $\sqrt{s} = 5.02$ TeV and p--Pb collisions at $\sqrt{s_{\rm{NN}}} = 5.02$ TeV. The statistical (systematic) uncertainties are shown as vertical lines (empty boxes). The uncertainties of the baseline are shown as solid boxes at $\Delta\varphi \sim$ -2 rad.}
\label{fig:AddDelphipPb412}
\end{figure}

\begin{figure}[!h]
\centering
\subfigure{
\includegraphics[scale=0.8]{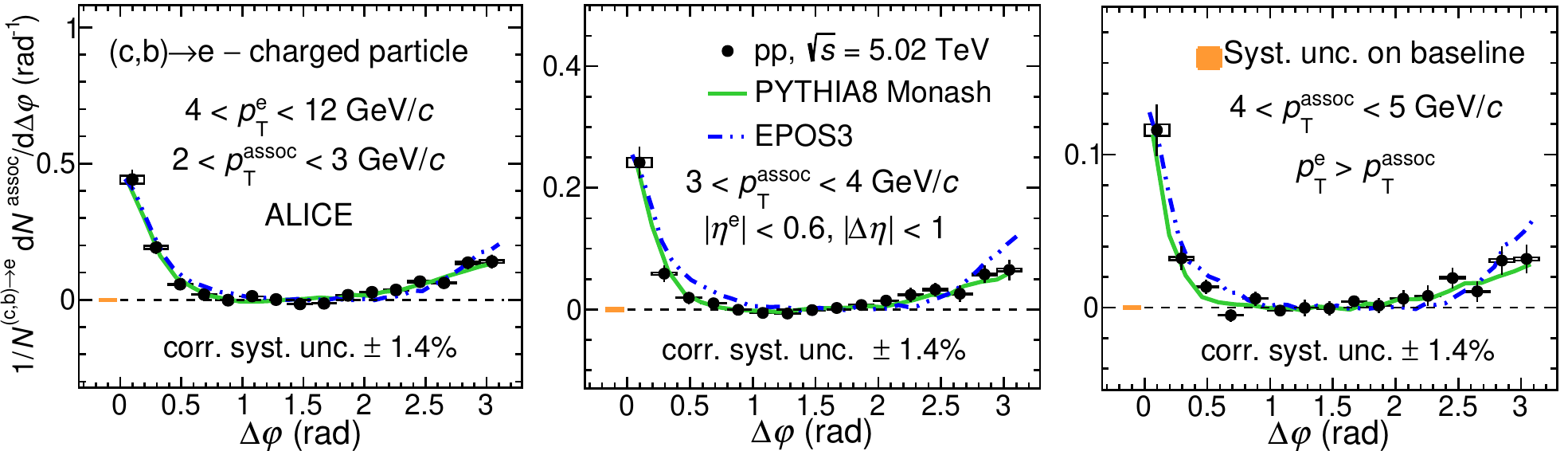}
}
\caption{Azimuthal-correlation distributions after baseline subtraction for $4 < \pt^{\rm e} < 12$ \GeVc ~in different $\pt^{\rm{assoc}}$ ranges compared with predictions from PYTHIA8 Monash and EPOS3 in pp collisions at $\sqrt{s} =$ 5.02 TeV. The statistical (systematic) uncertainties are shown as vertical lines (empty boxes). The uncertainties of the baseline are shown as solid boxes at $\dph \sim 0$ rad.}
\label{fig:AddMCDelphi_pp412}
\end{figure}

\begin{figure}[!h]
\centering
\subfigure{
\includegraphics[scale=0.8]{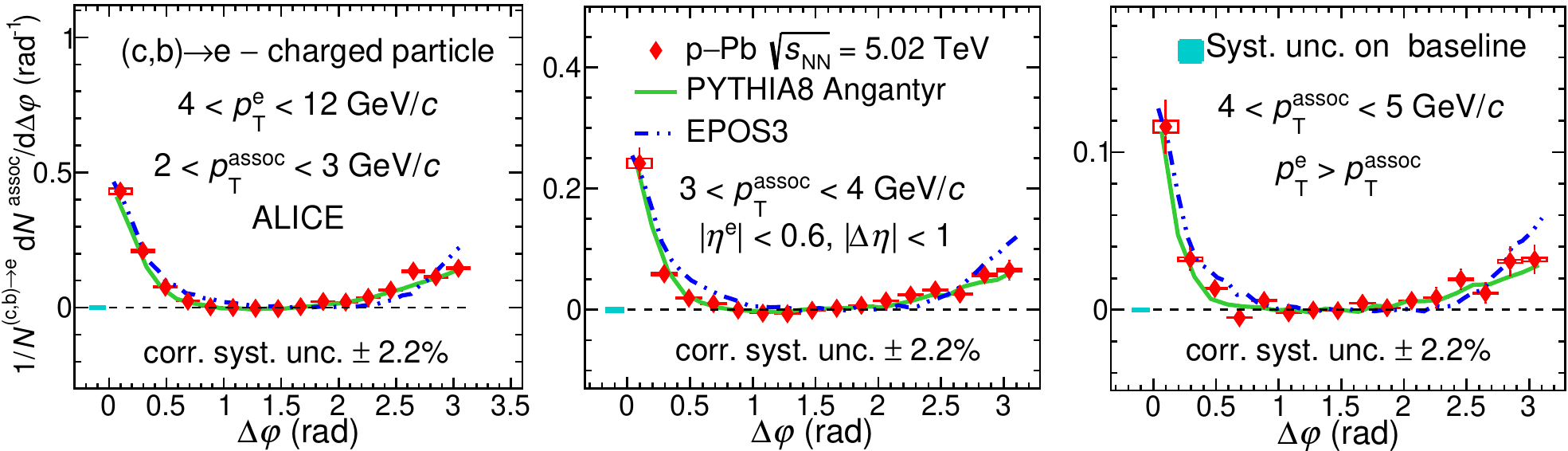}
}
\caption{Azimuthal-correlation distributions after baseline subtraction for $4 < \pt^{\rm e} < 12$ \GeVc ~in different $\pt^{\rm{assoc}}$ ranges compared with predictions from PYTHIA8 Angantyr and EPOS3 in p--Pb collisions at $\sqrt{s_{\rm{NN}}} =$ 5.02 TeV. The statistical (systematic) uncertainties are shown as vertical lines (empty boxes). The uncertainties of the baseline are shown as solid boxes at $\dph \sim 0$ rad.}
\label{fig:AddMCDelphipPb412}
\end{figure}

\begin{figure}[!h]
\centering
\subfigure{
\includegraphics[scale=0.8]{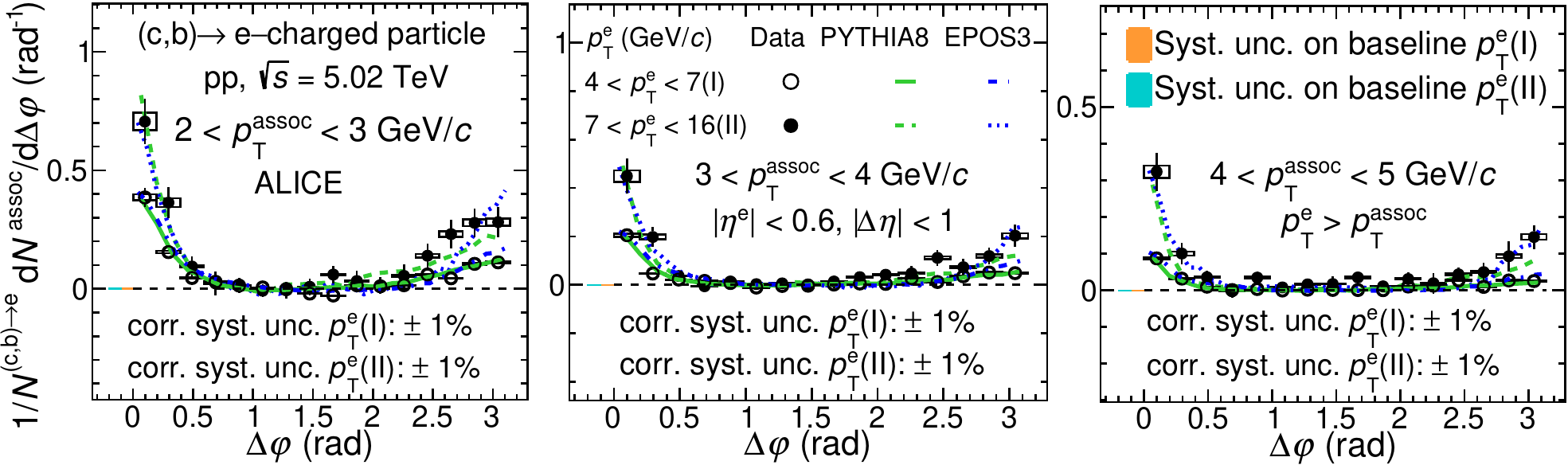}
}
\caption{Azimuthal-correlation distributions after baseline subtraction for two $\pt^{\rm e}$ intervals, $4 < \pt^{\rm e} < 7$ \GeVc and $7 < \pt^{\rm e} < 16$ \GeVc, and for different associated \pt ranges compared with predictions from PYTHIA8 Monash and EPOS3 in pp collisions at $\sqrt{s} =$ 5.02 TeV. The statistical (systematic) uncertainties are shown as vertical lines (empty boxes). The uncertainties of the baseline are shown as solid boxes at $\dph \sim 0$ rad.}
\label{fig:AddDelphi_pp_pT47_716}
\end{figure}

\begin{figure}[!h]
\centering
\subfigure{
\includegraphics[scale=0.8]{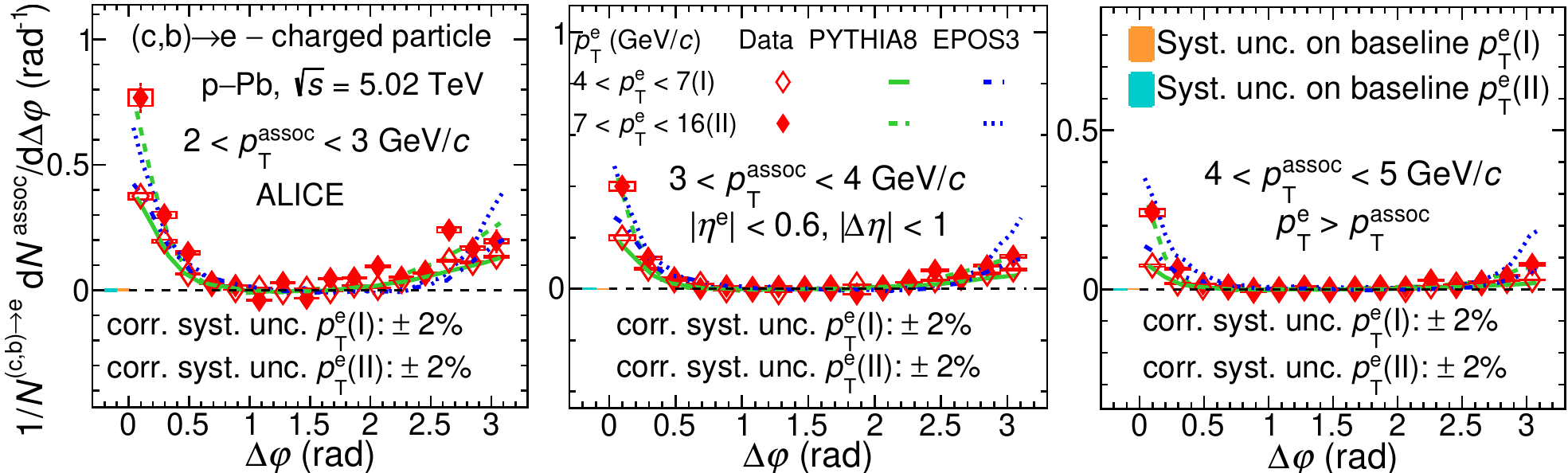}
}
\caption{Azimuthal-correlation distributions after baseline subtraction for two $\pt^{\rm e}$ intervals, $4 < \pt^{\rm e} < 7$ \GeVc and $7 < \pt^{\rm e} < 16$ \GeVc, and for different associated \pt ranges compared with predictions from PYTHIA8 Angantyr and EPOS3 in p--Pb collisions at $\sqrt{s_{\rm{NN}}} =$ 5.02 TeV. The statistical (systematic) uncertainties are shown as vertical lines (empty boxes). The uncertainties of the baseline are shown as solid boxes at $\dph \sim 0$ rad.}
\label{fig:AddMCDelphi_pPb_pT47_716}
\end{figure}

In this appendix, the azimuthal correlation distributions are shown for $\pt^{\rm{assoc}}$ ranges not presented in Sec.~\ref{sec:Results} along with comparisons to PYTHIA8 and EPOS3 predictions (Figs.~\ref{fig:AddDelphipPb412}, \ref{fig:AddMCDelphi_pp412}, \ref{fig:AddMCDelphipPb412}, \ref{fig:AddDelphi_pp_pT47_716}, \ref{fig:AddMCDelphi_pPb_pT47_716}).

\clearpage
\newpage

\section{The ALICE Collaboration}
\label{app:collab}
\begin{flushleft} 
\small

S.~Acharya\,\orcidlink{0000-0002-9213-5329}\,$^{\rm 125}$, 
D.~Adamov\'{a}\,\orcidlink{0000-0002-0504-7428}\,$^{\rm 86}$, 
A.~Adler$^{\rm 69}$, 
G.~Aglieri Rinella\,\orcidlink{0000-0002-9611-3696}\,$^{\rm 32}$, 
M.~Agnello\,\orcidlink{0000-0002-0760-5075}\,$^{\rm 29}$, 
N.~Agrawal\,\orcidlink{0000-0003-0348-9836}\,$^{\rm 50}$, 
Z.~Ahammed\,\orcidlink{0000-0001-5241-7412}\,$^{\rm 132}$, 
S.~Ahmad\,\orcidlink{0000-0003-0497-5705}\,$^{\rm 15}$, 
S.U.~Ahn\,\orcidlink{0000-0001-8847-489X}\,$^{\rm 70}$, 
I.~Ahuja\,\orcidlink{0000-0002-4417-1392}\,$^{\rm 37}$
% ...

\section*{Affiliation Notes}

$^{\rm I}$ Deceased\\
$^{\rm II}$ Also at: Max-Planck-Institut f\"{u}r Physik, Munich, Germany\\
$^{\rm III}$ Also at: Italian National Agency for New Technologies, Energy and Sustainable Economic Development (ENEA), Bologna, Italy\\
$^{\rm IV}$ Also at: Dipartimento DET del Politecnico di Torino, Turin, Italy\\
$^{\rm V}$ Also at: Department of Applied Physics, Aligarh Muslim University, Aligarh, India\\
$^{\rm VI}$ Also at: Institute of Theoretical Physics, University of Wroclaw, Poland\\
$^{\rm VII}$ Also at: An institution covered by a cooperation agreement with CERN\\

\section*{Collaboration Institutes}

$^{1}$ A.I. Alikhanyan National Science Laboratory (Yerevan Physics Institute) Foundation, Yerevan, Armenia\\
$^{2}$ AGH University of Science and Technology, Cracow, Poland\\
$^{3}$ Bogolyubov Institute for Theoretical Physics, National Academy of Sciences of Ukraine, Kiev, Ukraine\\
$^{4}$ Bose Institute, Department of Physics  and Centre for Astroparticle Physics and Space Science (CAPSS), Kolkata, India\\
$^{5}$ California Polytechnic State University, San Luis Obispo, California, United States\\
$^{6}$ Central China Normal University, Wuhan, China\\
$^{7}$ Centro de Aplicaciones Tecnol\'{o}gicas y Desarrollo Nuclear (CEADEN), Havana, Cuba\\
$^{8}$ Centro de Investigaci\'{o}n y de Estudios Avanzados (CINVESTAV), Mexico City and M\'{e}rida, Mexico\\
$^{9}$ Chicago State University, Chicago, Illinois, United States\\
$^{10}$ China Institute of Atomic Energy, Beijing, China\\
$^{11}$ Chungbuk National University, Cheongju, Republic of Korea\\
$^{12}$ Comenius University Bratislava, Faculty of Mathematics, Physics and Informatics, Bratislava, Slovak Republic\\
$^{13}$ COMSATS University Islamabad, Islamabad, Pakistan\\
$^{14}$ Creighton University, Omaha, Nebraska, United States\\
$^{15}$ Department of Physics, Aligarh Muslim University, Aligarh, India\\
$^{16}$ Department of Physics, Pusan National University, Pusan, Republic of Korea\\
$^{17}$ Department of Physics, Sejong University, Seoul, Republic of Korea\\
$^{18}$ Department of Physics, University of California, Berkeley, California, United States\\
$^{19}$ Department of Physics, University of Oslo, Oslo, Norway\\
$^{20}$ Department of Physics and Technology, University of Bergen, Bergen, Norway\\
$^{21}$ Dipartimento di Fisica, Universit\`{a} di Pavia, Pavia, Italy\\
$^{22}$ Dipartimento di Fisica dell'Universit\`{a} and Sezione INFN, Cagliari, Italy\\
$^{23}$ Dipartimento di Fisica dell'Universit\`{a} and Sezione INFN, Trieste, Italy\\
$^{24}$ Dipartimento di Fisica dell'Universit\`{a} and Sezione INFN, Turin, Italy\\
$^{25}$ Dipartimento di Fisica e Astronomia dell'Universit\`{a} and Sezione INFN, Bologna, Italy\\
$^{26}$ Dipartimento di Fisica e Astronomia dell'Universit\`{a} and Sezione INFN, Catania, Italy\\
$^{27}$ Dipartimento di Fisica e Astronomia dell'Universit\`{a} and Sezione INFN, Padova, Italy\\
$^{28}$ Dipartimento di Fisica `E.R.~Caianiello' dell'Universit\`{a} and Gruppo Collegato INFN, Salerno, Italy\\
$^{29}$ Dipartimento DISAT del Politecnico and Sezione INFN, Turin, Italy\\
$^{30}$ Dipartimento di Scienze MIFT, Universit\`{a} di Messina, Messina, Italy\\
$^{31}$ Dipartimento Interateneo di Fisica `M.~Merlin' and Sezione INFN, Bari, Italy\\
$^{32}$ European Organization for Nuclear Research (CERN), Geneva, Switzerland\\
$^{33}$ Faculty of Electrical Engineering, Mechanical Engineering and Naval Architecture, University of Split, Split, Croatia\\
$^{34}$ Faculty of Engineering and Science, Western Norway University of Applied Sciences, Bergen, Norway\\
$^{35}$ Faculty of Nuclear Sciences and Physical Engineering, Czech Technical University in Prague, Prague, Czech Republic\\
$^{36}$ Faculty of Physics, Sofia University, Sofia, Bulgaria\\
$^{37}$ Faculty of Science, P.J.~\v{S}af\'{a}rik University, Ko\v{s}ice, Slovak Republic\\
$^{38}$ Frankfurt Institute for Advanced Studies, Johann Wolfgang Goethe-Universit\"{a}t Frankfurt, Frankfurt, Germany\\
$^{39}$ Fudan University, Shanghai, China\\
$^{40}$ Gangneung-Wonju National University, Gangneung, Republic of Korea\\
$^{41}$ Gauhati University, Department of Physics, Guwahati, India\\
$^{42}$ Helmholtz-Institut f\"{u}r Strahlen- und Kernphysik, Rheinische Friedrich-Wilhelms-Universit\"{a}t Bonn, Bonn, Germany\\
$^{43}$ Helsinki Institute of Physics (HIP), Helsinki, Finland\\
$^{44}$ High Energy Physics Group,  Universidad Aut\'{o}noma de Puebla, Puebla, Mexico\\
$^{45}$ Horia Hulubei National Institute of Physics and Nuclear Engineering, Bucharest, Romania\\
$^{46}$ Indian Institute of Technology Bombay (IIT), Mumbai, India\\
$^{47}$ Indian Institute of Technology Indore, Indore, India\\
$^{48}$ INFN, Laboratori Nazionali di Frascati, Frascati, Italy\\
$^{49}$ INFN, Sezione di Bari, Bari, Italy\\
$^{50}$ INFN, Sezione di Bologna, Bologna, Italy\\
$^{51}$ INFN, Sezione di Cagliari, Cagliari, Italy\\
$^{52}$ INFN, Sezione di Catania, Catania, Italy\\
$^{53}$ INFN, Sezione di Padova, Padova, Italy\\
$^{54}$ INFN, Sezione di Pavia, Pavia, Italy\\
$^{55}$ INFN, Sezione di Torino, Turin, Italy\\
$^{56}$ INFN, Sezione di Trieste, Trieste, Italy\\
$^{57}$ Inha University, Incheon, Republic of Korea\\
$^{58}$ Institute for Gravitational and Subatomic Physics (GRASP), Utrecht University/Nikhef, Utrecht, Netherlands\\
$^{59}$ Institute of Experimental Physics, Slovak Academy of Sciences, Ko\v{s}ice, Slovak Republic\\
$^{60}$ Institute of Physics, Homi Bhabha National Institute, Bhubaneswar, India\\
$^{61}$ Institute of Physics of the Czech Academy of Sciences, Prague, Czech Republic\\
$^{62}$ Institute of Space Science (ISS), Bucharest, Romania\\
$^{63}$ Institut f\"{u}r Kernphysik, Johann Wolfgang Goethe-Universit\"{a}t Frankfurt, Frankfurt, Germany\\
$^{64}$ Instituto de Ciencias Nucleares, Universidad Nacional Aut\'{o}noma de M\'{e}xico, Mexico City, Mexico\\
$^{65}$ Instituto de F\'{i}sica, Universidade Federal do Rio Grande do Sul (UFRGS), Porto Alegre, Brazil\\
$^{66}$ Instituto de F\'{\i}sica, Universidad Nacional Aut\'{o}noma de M\'{e}xico, Mexico City, Mexico\\
$^{67}$ iThemba LABS, National Research Foundation, Somerset West, South Africa\\
$^{68}$ Jeonbuk National University, Jeonju, Republic of Korea\\
$^{69}$ Johann-Wolfgang-Goethe Universit\"{a}t Frankfurt Institut f\"{u}r Informatik, Fachbereich Informatik und Mathematik, Frankfurt, Germany\\
$^{70}$ Korea Institute of Science and Technology Information, Daejeon, Republic of Korea\\
$^{71}$ KTO Karatay University, Konya, Turkey\\
$^{72}$ Laboratoire de Physique des 2 Infinis, Ir\`{e}ne Joliot-Curie, Orsay, France\\
$^{73}$ Laboratoire de Physique Subatomique et de Cosmologie, Universit\'{e} Grenoble-Alpes, CNRS-IN2P3, Grenoble, France\\
$^{74}$ Lawrence Berkeley National Laboratory, Berkeley, California, United States\\
$^{75}$ Lund University Department of Physics, Division of Particle Physics, Lund, Sweden\\
$^{76}$ Nagasaki Institute of Applied Science, Nagasaki, Japan\\
$^{77}$ Nara Women{'}s University (NWU), Nara, Japan\\
$^{78}$ National and Kapodistrian University of Athens, School of Science, Department of Physics , Athens, Greece\\
$^{79}$ National Centre for Nuclear Research, Warsaw, Poland\\
$^{80}$ National Institute of Science Education and Research, Homi Bhabha National Institute, Jatni, India\\
$^{81}$ National Nuclear Research Center, Baku, Azerbaijan\\
$^{82}$ National Research and Innovation Agency - BRIN, Jakarta, Indonesia\\
$^{83}$ Niels Bohr Institute, University of Copenhagen, Copenhagen, Denmark\\
$^{84}$ Nikhef, National institute for subatomic physics, Amsterdam, Netherlands\\
$^{85}$ Nuclear Physics Group, STFC Daresbury Laboratory, Daresbury, United Kingdom\\
$^{86}$ Nuclear Physics Institute of the Czech Academy of Sciences, Husinec-\v{R}e\v{z}, Czech Republic\\
$^{87}$ Oak Ridge National Laboratory, Oak Ridge, Tennessee, United States\\
$^{88}$ Ohio State University, Columbus, Ohio, United States\\
$^{89}$ Physics department, Faculty of science, University of Zagreb, Zagreb, Croatia\\
$^{90}$ Physics Department, Panjab University, Chandigarh, India\\
$^{91}$ Physics Department, University of Jammu, Jammu, India\\
$^{92}$ Physics Program and International Institute for Sustainability with Knotted Chiral Meta Matter (SKCM2), Hiroshima University, Hiroshima, Japan\\
$^{93}$ Physikalisches Institut, Eberhard-Karls-Universit\"{a}t T\"{u}bingen, T\"{u}bingen, Germany\\
$^{94}$ Physikalisches Institut, Ruprecht-Karls-Universit\"{a}t Heidelberg, Heidelberg, Germany\\
$^{95}$ Physik Department, Technische Universit\"{a}t M\"{u}nchen, Munich, Germany\\
$^{96}$ Politecnico di Bari and Sezione INFN, Bari, Italy\\
$^{97}$ Research Division and ExtreMe Matter Institute EMMI, GSI Helmholtzzentrum f\"ur Schwerionenforschung GmbH, Darmstadt, Germany\\
$^{98}$ Saga University, Saga, Japan\\
$^{99}$ Saha Institute of Nuclear Physics, Homi Bhabha National Institute, Kolkata, India\\
$^{100}$ School of Physics and Astronomy, University of Birmingham, Birmingham, United Kingdom\\
$^{101}$ Secci\'{o}n F\'{\i}sica, Departamento de Ciencias, Pontificia Universidad Cat\'{o}lica del Per\'{u}, Lima, Peru\\
$^{102}$ Stefan Meyer Institut f\"{u}r Subatomare Physik (SMI), Vienna, Austria\\
$^{103}$ SUBATECH, IMT Atlantique, Nantes Universit\'{e}, CNRS-IN2P3, Nantes, France\\
$^{104}$ Sungkyunkwan University, Suwon City, Republic of Korea\\
$^{105}$ Suranaree University of Technology, Nakhon Ratchasima, Thailand\\
$^{106}$ Technical University of Ko\v{s}ice, Ko\v{s}ice, Slovak Republic\\
$^{107}$ The Henryk Niewodniczanski Institute of Nuclear Physics, Polish Academy of Sciences, Cracow, Poland\\
$^{108}$ The University of Texas at Austin, Austin, Texas, United States\\
$^{109}$ Universidad Aut\'{o}noma de Sinaloa, Culiac\'{a}n, Mexico\\
$^{110}$ Universidade de S\~{a}o Paulo (USP), S\~{a}o Paulo, Brazil\\
$^{111}$ Universidade Estadual de Campinas (UNICAMP), Campinas, Brazil\\
$^{112}$ Universidade Federal do ABC, Santo Andre, Brazil\\
$^{113}$ University of Cape Town, Cape Town, South Africa\\
$^{114}$ University of Houston, Houston, Texas, United States\\
$^{115}$ University of Jyv\"{a}skyl\"{a}, Jyv\"{a}skyl\"{a}, Finland\\
$^{116}$ University of Kansas, Lawrence, Kansas, United States\\
$^{117}$ University of Liverpool, Liverpool, United Kingdom\\
$^{118}$ University of Science and Technology of China, Hefei, China\\
$^{119}$ University of South-Eastern Norway, Kongsberg, Norway\\
$^{120}$ University of Tennessee, Knoxville, Tennessee, United States\\
$^{121}$ University of the Witwatersrand, Johannesburg, South Africa\\
$^{122}$ University of Tokyo, Tokyo, Japan\\
$^{123}$ University of Tsukuba, Tsukuba, Japan\\
$^{124}$ University Politehnica of Bucharest, Bucharest, Romania\\
$^{125}$ Universit\'{e} Clermont Auvergne, CNRS/IN2P3, LPC, Clermont-Ferrand, France\\
$^{126}$ Universit\'{e} de Lyon, CNRS/IN2P3, Institut de Physique des 2 Infinis de Lyon, Lyon, France\\
$^{127}$ Universit\'{e} de Strasbourg, CNRS, IPHC UMR 7178, F-67000 Strasbourg, France, Strasbourg, France\\
$^{128}$ Universit\'{e} Paris-Saclay Centre d'Etudes de Saclay (CEA), IRFU, D\'{e}partment de Physique Nucl\'{e}aire (DPhN), Saclay, France\\
$^{129}$ Universit\`{a} degli Studi di Foggia, Foggia, Italy\\
$^{130}$ Universit\`{a} del Piemonte Orientale, Vercelli, Italy\\
$^{131}$ Universit\`{a} di Brescia, Brescia, Italy\\
$^{132}$ Variable Energy Cyclotron Centre, Homi Bhabha National Institute, Kolkata, India\\
$^{133}$ Warsaw University of Technology, Warsaw, Poland\\
$^{134}$ Wayne State University, Detroit, Michigan, United States\\
$^{135}$ Westf\"{a}lische Wilhelms-Universit\"{a}t M\"{u}nster, Institut f\"{u}r Kernphysik, M\"{u}nster, Germany\\
$^{136}$ Wigner Research Centre for Physics, Budapest, Hungary\\
$^{137}$ Yale University, New Haven, Connecticut, United States\\
$^{138}$ Yonsei University, Seoul, Republic of Korea\\
$^{139}$  Zentrum  f\"{u}r Technologie und Transfer (ZTT), Worms, Germany\\
$^{140}$ Affiliated with an institute covered by a cooperation agreement with CERN\\
$^{141}$ Affiliated with an international laboratory covered by a cooperation agreement with CERN.\\

\end{flushleft} 
\end{document}